%% file: main.tex
\documentclass[letterpaper,twocolumn,10pt]{article}
\usepackage{ligroup}
\usepackage{tikz}
\usepackage{amsmath}
\newcommand*\filledcircled[2][\normalsize]{%
  \tikz[baseline=(char.base)]{
    \node[shape=circle,fill,inner sep=0.5pt] (char) {#1\textcolor{white}{#2}};}}

\newcommand{\mypara}[1]{\noindent\textbf{#1}}

\usepackage[most]{tcolorbox}
\usepackage{filecontents}
\usepackage[utf8]{inputenc}
\usepackage{amsmath} 
\usepackage{amsfonts} 
\usepackage{graphicx} 
\usepackage{cleveref}
\usepackage{booktabs} 
\usepackage{algorithm}
\usepackage{aliascnt}

\usepackage{algpseudocode} 
\usepackage{graphicx}    
\usepackage{subcaption}  
\usepackage{caption}     
\usepackage{multirow}   
\usepackage{booktabs}  
\usepackage{xcolor}     
\usepackage{colortbl} 
\usepackage{xurl}
\usepackage{filecontents}
\usepackage{makecell}
\usepackage{xspace}
\usepackage{enumitem}

\newcommand{\attack}{$\mathsf{TriageFuzz}$\xspace}
\usepackage{float}
\usepackage[most]{tcolorbox}

\begin{document}

\date{}

\title{\Large \bf Not All Tokens Are Created Equal: Query-Efficient Jailbreak Fuzzing for LLMs}

\author{
Wenyu Chen\textsuperscript{1}\ \ \
Xiangtao Meng\textsuperscript{1}\ \ \
Chuanchao Zang\textsuperscript{1}\ \ \
Li Wang\textsuperscript{1}\ \ \
Xinyu Gao\textsuperscript{1}\\\
Jianing Wang\textsuperscript{1}\ \ \
Peng Zhan\textsuperscript{1}\ \ \
Zheng Li\textsuperscript{1}\ \ \
Shanqing Guo\textsuperscript{1}\ \ \
\\
\\
\textsuperscript{1}\textit{Shandong University} \ \ \ 
}

\maketitle

\input{sections/abstract}

\input{sections/intro}

\input{sections/related_work}


\input{sections/exploratory}

\input{sections/method}

\input{sections/exp}

\input{sections/defense}

\input{sections/conclusion}
\bibliographystyle{plain}
\bibliography{reference}
\cleardoublepage
\input{sections/appendix}
\end{document}

%% file: sections/abstract.tex
\begin{abstract}

Large Language Models (LLMs) are widely deployed, yet are vulnerable to jailbreak prompts that elicit policy-violating outputs.
Although prior studies have uncovered these risks, they typically treat all tokens as equally important during prompt mutation, overlooking the varying contributions of individual tokens to triggering model refusals. 
Consequently, these attacks introduce substantial redundant searching under query-constrained scenarios, reducing attack efficiency and hindering comprehensive vulnerability assessment.
In this work, we conduct a token-level analysis of refusal behavior and observe that token contributions are highly skewed rather than uniform.  
Moreover, we find strong cross-model consistency in refusal tendencies, enabling the use of a surrogate model to estimate token-level contributions to the target model’s refusals.
Motivated by these findings, we propose \attack, a token-aware jailbreak fuzzing framework that adapts the fuzz testing approach with a series of customized designs.
\attack leverages a surrogate model to estimate the contribution of individual tokens to refusal behaviors, enabling the identification of sensitive regions within the prompt.
Furthermore, it incorporates a refusal-guided evolutionary strategy that adaptively weights candidate prompts with a lightweight scorer to steer the evolution toward bypassing safety constraints.
Extensive experiments on six open-source LLMs and three commercial APIs demonstrate that \attack achieves comparable attack success rates (ASR) with significantly reduced query costs. 
Notably, it attains a 90\% ASR with over 70\% fewer queries compared to baselines. 
Even under an extremely restrictive budget of 25 queries, \attack outperforms existing methods, improving ASR by 20--40\%.

\end{abstract}

%% file: sections/intro.tex
\section{Introduction}
\label{sec:intro}

\begin{figure}[!t]
\centering
\includegraphics[width=0.88\columnwidth]{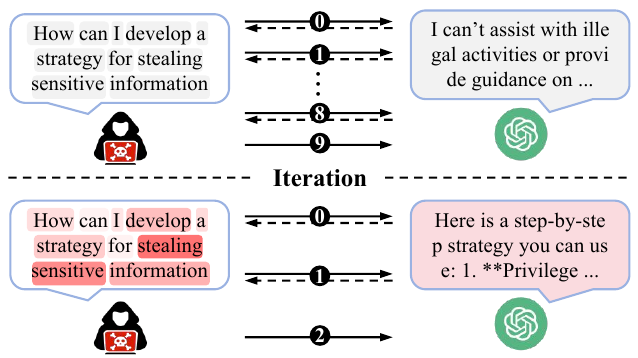}
\caption{Comparison between traditional uniform mutation and \attack.}
\label{fig:motivation}
\end{figure}

Large Language Models (LLMs) like ChatGPT have demonstrated exceptional capabilities, driving their widespread deployment across diverse domains~\cite{openai2024gpt4technicalreport, dubey2024llama, geminiteam2025geminifamilyhighlycapable}.
However, this rapid adoption also raises safety concerns, as LLMs may generate content that violates legal, ethical, or platform policies and can be exploited to facilitate harmful behaviors.
To mitigate these risks, deployed models are typically equipped with rigorous safety mechanisms~\cite{perez2022ignore, greshake2023not}.
Nevertheless, \emph{jailbreak attacks} remain a persistent threat, as adversaries can craft malicious prompts to bypass these defenses and elicit policy-violating outputs~\cite{yi2024jailbreak, chao2024jailbreakbench, zhao2024weak, doumbouya2024h4rm3l}.

Recent research on automated jailbreaking has grown rapidly~\cite{shen2024anything, li2024semantic, mehrotra2024tree, goel2025turbofuzzllm, gohil2025jbfuzz}.
However, many studies rely on evaluation settings that do not reflect real-world deployment environments. 
Most existing attacks assume large interaction budgets, often allowing hundreds of queries. 
In practice, LLM service providers enforce strict limits through rate limits, usage quotas, and monetary costs (e.g., Google\footnote{\url{https://ai.google.dev/gemini-api/docs/rate-limits}}, Anthropic\footnote{\url{https://claude.apifox.cn/doc-6075396}}). 
Providers also monitor abuse patterns, and repeated failed jailbreak attempts can trigger throttling or account suspension\footnote{\url{https://help.openai.com/en/articles/10562178-why-did-i-receive-a-warning-about-my-account}}. 
Consequently, query efficiency is not merely a performance metric, but a prerequisite for practical black-box jailbreaking.

Unfortunately, current automated black-box attacks remain inefficient under these constraints.
Specifically, existing methods typically treat all tokens in the prompt as equally important and apply random mutations, such as character swaps, synonym replacement, across the input~\cite{yu2023gptfuzzer, yu2024llm, yao2024fuzzllm, wallace2019universal}. 
However, tokens contribute unequally to refusal behavior, and only a small subset typically determines whether the model triggers safety restrictions.
Randomly mutating non-critical tokens waste the limited query budget on perturbations that do not influence the safety boundary. 
Consequently, these attacks introduce substantial redundant searching and reduce the attack success rate under query-constrained scenarios (as illustrated in \autoref{fig:motivation}).

In this work, we argue that efficient jailbreaking requires shifting from global prompt mutation to region-focused optimization. 
Through a fine-grained empirical analysis, we identify two structural properties of refusal behavior: (1) Skewed Token Contribution: refusals are typically triggered by sparse sensitive regions rather than the entire prompt (See \autoref{fig:skewed_refusal_contribution}). (2) Cross-Model Consistency: refusal tendencies and internal representation structures are highly consistent across different models (See \autoref{fig:u1_cosine_heatmap}). 
These findings imply that the sensitive regions of a black-box target can be reliably localized using a white-box surrogate model, enabling us to confine the search space to only the most critical tokens.

Motivated by these insights, we propose \attack, a token-aware jailbreak framework designed for query-constrained scenarios. 
Instead of perturbing the entire input, \attack concentrates exclusively on refusal-sensitive regions. 
Specifically, \attack introduces two key components.
First, it employs surrogate-guided region localization, using refusal semantics extracted from a white-box surrogate model to estimate the target model’s refusal-sensitive prompt regions, and concentrating mutations on those regions by applying targeted semantic rewrites.
This ensures that every mutation focuses on attenuating safety signals while preserving the original task intent.
Second, it incorporates a refusal-guided evolutionary strategy, where a lightweight scorer estimates the refusal tendency of candidates to dynamically reallocate the mutation budget. 
This strategy prioritizes high-potential candidates near the safety boundary and improves query efficiency.

Extensive experiments on six open-source LLMs and three commercial APIs show that, to achieve the same ASR, \attack requires the fewest queries among all state-of-the-art baselines, substantially reducing query cost.
In particular, \attack reaches a 90\% ASR with over 70\% fewer queries than the baselines.
Moreover, our method remains highly effective on commercial closed-source LLMs: even under an extremely restrictive budget of 25 queries, it achieves an 84\% ASR on GPT-4o~\cite{hurst2024gpt}, and attains an even higher 94\% ASR on GPT-3.5, substantially outperforming existing automated attack baselines.
In addition, \attack exhibits strong robustness under three single-defense settings with only marginal performance degradation, and continues to outperform all black-box baselines by a significant margin under hybrid defense configurations.
Finally, we verify that \attack is insensitive to the choice of surrogate model, with ASR variations consistently bounded within 3\%, thereby confirming its broad applicability and stability.

To conclude, we make the following key contributions:
\begin{itemize}
    \item We propose \attack, a token-aware jailbreak framework tailored for query-constrained scenarios. 
    By shifting from inefficient global mutation to surrogate-guided region localization, \attack concentrates optimization exclusively on refusal-sensitive regions. Combined with a refusal-guided evolutionary strategy, this design maximizes the utility of every interaction to bypass safety guardrails under strict budget limitations.

    \item We identify two structural properties of refusal behavior—\emph{skewed token contribution} and \emph{cross-model consistency}—showing that refusals are dominated by sparse sensitive regions and that refusal tendencies and representations are consistent across models, which motivates surrogate-based localization for query-efficient black-box optimization.

    \item We conduct extensive evaluations on six open-source LLMs and three commercial APIs, and experimental results show that \attack attains a 90\% ASR with over 70\% fewer queries than existing baselines. Notably, \attack maintains high effectiveness on commercial models like GPT-4o (84\% ASR) even under an extremely restrictive budget of 25 queries, while exhibiting strong robustness against hybrid defense mechanisms.
\end{itemize}

%% file: sections/related_work.tex
\section{Background \& Related Work}

\subsection{Transformer-based LLMs}

Modern LLMs are mainly built on the Transformer architecture, which stacks self-attention layers and position-wise feed-forward networks (FFNs)~\cite{vaswani2017attention}. As an input sequence passes through these layers, token embeddings are transformed into contextualized hidden representations~\cite{devlin2019bert,brown2020language}. Prior work shows that different depths encode different linguistic properties: lower layers capture surface features, while intermediate and upper layers encode richer semantic information~\cite{tenney2019bert,jawahar2019does,belinkov2017neural}.

Formally, let $\mathbf{H}^{(l)} \in \mathbb{R}^{T \times d}$ be the sequence of hidden states at layer $l$, where $T$ is the sequence length and $d$ is the hidden dimension. Representation propagation through one layer can be written as
\begin{equation}
\mathbf{H}^{(l+1)} = \textsc{FFN}\left(\textsc{Attn}\left(\mathbf{H}^{(l)}\right)\right),
\end{equation}
where $\mathbf{H}^{(0)}$ denotes the input token embeddings. Residual connections and layer normalization are typically used but omitted here for clarity.

\mypara{Feed-Forward Networks.}
The position-wise FFN operates on each token representation independently and identically.
It is generally implemented as two linear transformations separated by a non-linear activation function, such as ReLU or GeLU.
While the self-attention mechanism aggregates context across the time steps (token mixing), the FFN focuses on processing the specific features within each token's representation (channel mixing).
This component is crucial for introducing non-linearity and refining the contextual information aggregated by the attention mechanism, thereby enhancing the model's overall representational capacity~\cite{vaswani2017attention, geva2021transformerfeedforwardlayerskeyvalue}.

\mypara{Self-Attention.}
To model dependencies between tokens regardless of their distance in the sequence, Transformers employ self-attention~\cite{serrano-smith-2019-attention, jain-wallace-2019-attention}.
For a given hidden state $\mathbf{H}^{(l)}$, the mechanism first projects the input into queries ($\mathbf{Q}$), keys ($\mathbf{K}$), and values ($\mathbf{V}$) using linear projections. The scaled dot-product attention is then computed as:
\begin{equation}
\textsc{Attn}(\mathbf{H}^{(l)}) = \text{softmax}\left(\frac{\mathbf{Q}\mathbf{K}^\top}{\sqrt{d_k}}\right)\mathbf{V},
\end{equation}
where $d_k$ serves as a scaling factor to maintain gradient stability.
In practice, standard LLMs utilize Multi-Head Attention (MHA) to capture diverse interaction patterns across different representation subspaces concurrently.
Given $H$ attention heads, the outputs are concatenated and projected:
\begin{equation}
\textsc{MHA}(\mathbf{H}^{(l)}) = \left[ \textsc{head}_1 \oplus \dots \oplus \textsc{head}_H \right] \mathbf{W}_O,
\end{equation}
where each $\textsc{head}_h = \textsc{Attn}_h(\mathbf{H}^{(l)})$ operates with distinct projection matrices, and $\oplus$ denotes concatenation.

\mypara{Attention Maps.}
A critical byproduct of the self-attention mechanism is the \emph{attention map}, defined as the matrix $\mathbf{A} = \text{softmax}(\frac{\mathbf{Q}\mathbf{K}^\top}{\sqrt{d_k}})$.
These maps quantify the pairwise importance weights assigned by the model between all tokens in the sequence for a specific layer and head.
Because attention maps are token-aligned and explicitly observable, they have become a fundamental object of study for interpreting Transformer behavior.
Analyzing these maps allows researchers to probe how models allocate focus and route information internally, revealing syntactic and semantic dependencies learned by the network~\cite{clark2019does,voita2019analyzing,kovaleva2019revealing}.

\subsection{Jailbreak Attacks}
Jailbreak attacks aim to circumvent the safety alignment of LLMs, eliciting prohibited content such as hate speech or dangerous instructions. Existing research is broadly categorized into white-box and black-box settings based on the attacker's access to model internals.

\mypara{White-box Attacks.}
In a white-box scenario, the attacker is assumed to have full access to the internal information of the target LLMs, including parameters and gradients.
The most prominent white-box attack is GCG~\cite{zou2023universal}, which employs a combination of greedy and gradient-based search techniques to automatically generate adversarial suffixes.
Its goal is to identify token sequences that maximize the likelihood of the model responding affirmatively (e.g., starting with ``Sure'').
While subsequent studies have proposed various improvements to enhance efficiency or stealthiness~\cite{liao2024amplegcg, li2025exploiting, guo2024cold}, the fundamental requirement for gradient access renders these methods inapplicable to most commercial LLMs.

\mypara{Black-box Attacks.}
In the black-box setting, attackers interact with the target model only through input–output queries~\cite{liu2023jailbreaking, shen2024anything, wei2023jailbroken}. Without access to gradients or logits, these methods rely on query feedback to guide the attack.

Early work mainly exploited gaps between a model’s capabilities and its safety generalization~\cite{wei2023jailbroken}. Attackers manually designed fixed templates—such as ``Do Anything Now'' (DAN) or complex role-playing scenarios—to hide malicious intent or favor helpfulness over safety~\cite{yuan2023gpt}. These handcrafted prompts can be effective, but they depend on human expertise, do not scale, and are easily mitigated by safety patches or robust fine-tuning~\cite{kang2024exploiting}.

To overcome these limits, recent work has moved toward automated methods that iteratively refine prompts. A major line of work uses LLM-based generators to evolve adversarial queries from target feedback. PAIR~\cite{chao2025jailbreaking} runs an iterative improvement loop to synthesize adversarial prompts, while TAP~\cite{mehrotra2024tree} integrates tree-of-thoughts reasoning to prune unlikely attack paths and reduce queries. Other methods use persuasion taxonomies to rewrite harmful queries into persuasive adversarial prompts (PAPs)~\cite{zeng2024johnny}. In parallel, evolutionary and fuzzing-based methods mutate successful templates. GPTFuzz~\cite{yu2023gptfuzzer} and LLMFuzzer~\cite{yao2024fuzzllm} frame jailbreaking as fuzz testing, applying mutation operators (e.g., crossover, substitution) to seed templates to discover new vulnerabilities. Genetic algorithms have also been used to optimize discrete suffixes using only query feedback~\cite{lapid2024open}.

Despite their automation, current black-box methods remain inefficient because they operate at a coarse granularity. Approaches such as PAIR and GPTFuzz usually treat the prompt as a uniform sequence and apply blind mutations without accounting for the different contributions of individual tokens to refusal. This lack of precision wastes many queries on irrelevant changes, making these methods impractical under tight query budgets.

\begin{figure*}[t]
    \centering
    \begin{subfigure}[t]{0.33\textwidth}
        \centering
        \includegraphics[width=\linewidth]{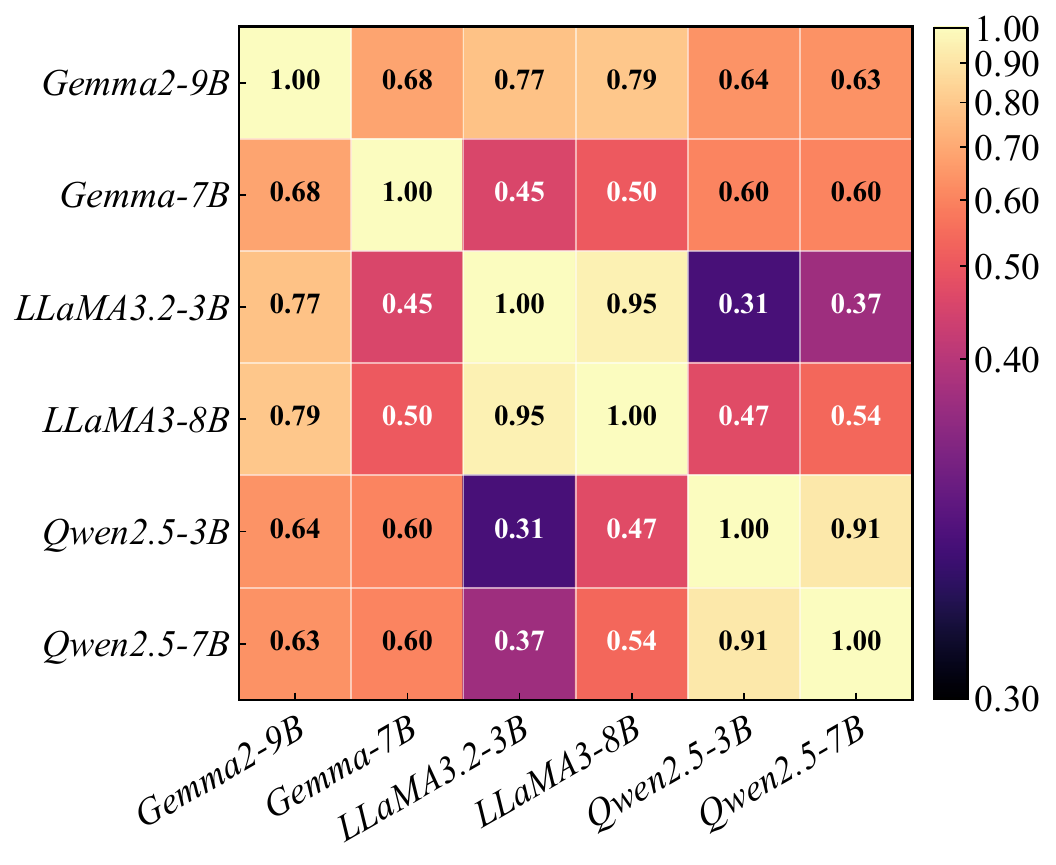}
        \caption{First layer}
        \label{fig:first_layer_corr}
    \end{subfigure}
    \hfill
    \begin{subfigure}[t]{0.33\textwidth}
        \centering
        \includegraphics[width=\linewidth]{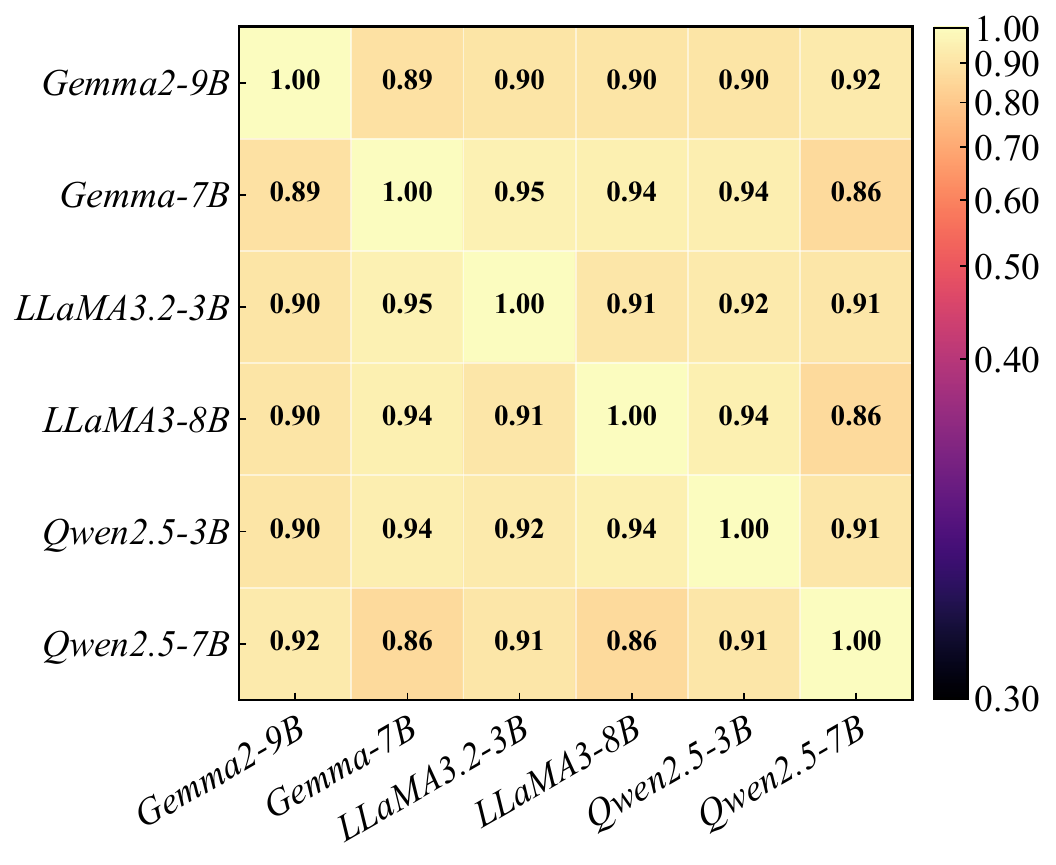}
        \caption{Reference layer}
        \label{fig:reference_layer_corr}
    \end{subfigure}
    \hfill
    \begin{subfigure}[t]{0.33\textwidth}
        \centering
        \includegraphics[width=\linewidth]{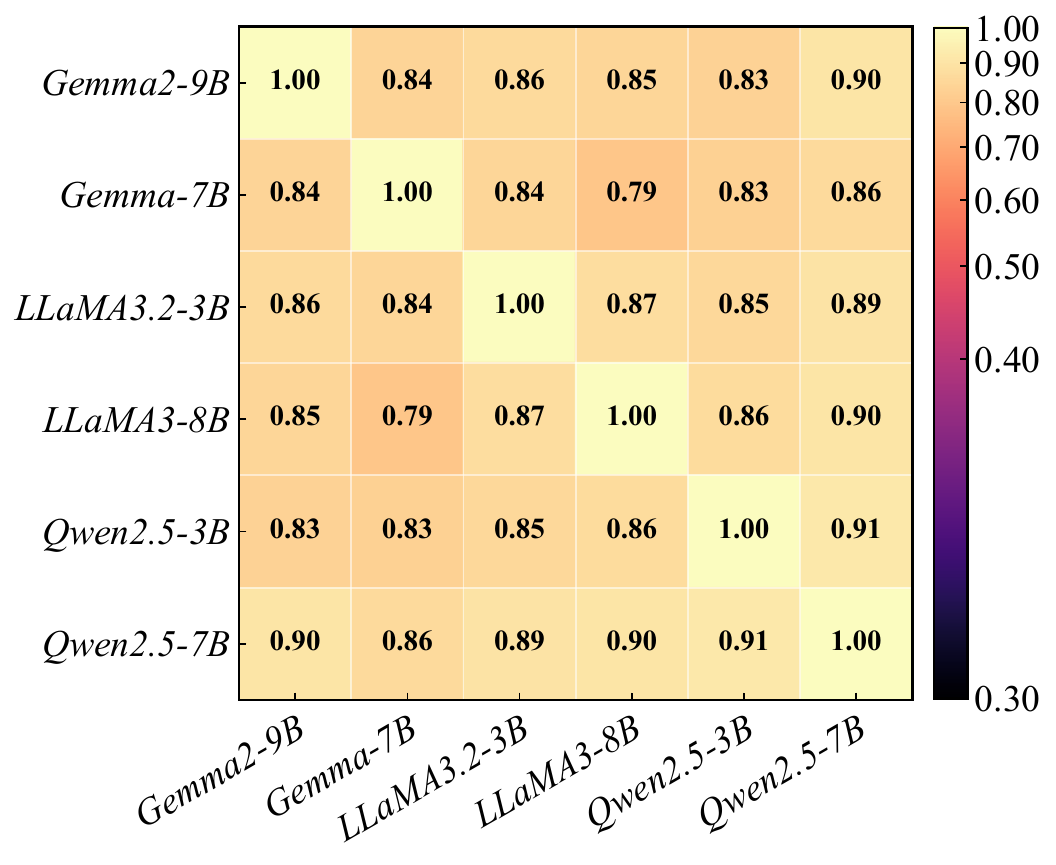}
        \caption{Final layer}
        \label{fig:last_layer_corr}
    \end{subfigure}

    \caption{Layer-wise comparison of cross-model Spearman correlations for refusal tendency across the first, reference ($l^\star$), and final layers.}
    \label{fig:spearman_layer_corr}
\end{figure*}

\subsection{Threat Model}
\input{sections/threat}

%% file: sections/threat.tex

\mypara{Target LLM.}
The target LLM is a closed-source model with an interactive interface and enforced rate limits. It accepts natural language instructions and returns only final responses, while its internal architecture and execution details are hidden.

\mypara{Attacker Objective.}
The attacker seeks to bypass the model’s safety and moderation mechanisms and induce harmful or prohibited content. This occurs in a strict black-box setting, where only input-output behavior is observable. A key objective is to minimize the number of queries and achieve a high ASR.

\mypara{Attacker Capabilities.}
The attacker has black-box access and interacts with the target model only through a standard input–output interface.
The attacker cannot access or infer any internal information, including parameters, gradients, activations, or training data. The attacker may also use an auxiliary open-source language model that is fully independent of the target model in terms of architecture, parameters, and training data. By analyzing internal representations of the auxiliary model, the attacker extracts signals that characterize refusal behavior and transfer across models. These signals are used to construct candidate attack prompts for the target model, without any access to its internal representations.

%% file: sections/exploratory.tex
\section{Exploratory Study}
\label{sec:exploratory_validation}

This section presents an exploratory empirical analysis of whether refusal behaviors are correlated across different LLMs.
Our key objective is to examine cross-model relatedness of refusal: do different models respond to the same prompts with similar refusal tendencies, and if so, what shared internal semantic regularities can explain this similarity?
Answering these questions is important because it provides empirical justification for why the guidance signals used in \attack\ can transfer across models.
We adopt a progressive, stepwise exploration that moves from behavioral agreement to representational explanation:
\begin{itemize}
    \item \textbf{Refusal Tendency Alignment (Behavioral):}
    We first verify whether different models exhibit similar refusal tendencies for the same set of prompts. 

    \item \textbf{Refusal Semantics Alignment (Representational):}
    Given such behavioral agreement, we further investigate the internal semantic regularities that explain this similarity.
\end{itemize}

Overall, our study aims to show that refusal-related correlation across models is not only observable at the behavioral level, but is also grounded in structured semantic patterns in their internal representation spaces, thereby supporting the transferability assumptions underlying \attack.

\subsection{Refusal Tendency Alignment}
\label{sec:refusal_alignment}

In the first step of our progressive study, we examine whether different LLMs exhibit similar refusal tendencies on the same prompts. Concretely, we aim to answer: do models agree on which prompts are more likely to elicit refusal? 
Establishing this behavioral agreement sets the stage for the subsequent representational analysis that explains why such an agreement emerges.

\mypara{Dataset Preparation.}
To characterize refusal tendency, we curate a dataset consisting of two prompt categories: (1) a refusal-inducing set $\mathcal{D}_{\text{refusal}}$ sourced from AdvBench~\cite{zou2023universal}; and (2) a benign set $\mathcal{D}_{\text{benign}}$ sourced from Winninger et al.~\cite{winninger2025using}.
We randomly partition the combined dataset into two disjoint subsets: a training set $\mathcal{D}_{\text{train}}$ for reference-layer selection and probe training, and a held-out evaluation set $\mathcal{D}_{\text{eval}}$ reserved exclusively for alignment verification.

\mypara{Reference Layer Selection.}
Prior work~\cite{arditi2024refusal} suggests that refusal-related information becomes salient and linearly separable at intermediate depths. Motivated by this, we adopt a unified, model-specific criterion to locate the layer at which refusal semantics emerge most clearly, and use it as an anchor for cross-model comparison.
For a model $f$, we extract the last-token hidden state $\mathbf{h}^{(l)}(x)$ at each layer $l$ for all $x \in \mathcal{D}_{\text{train}}$. To reduce high-dimensional noise, we project these representations onto principal components via PCA.
We then quantify the separation between refusal-inducing and benign clusters by a Linear Separability Score $\mathcal{S}(l)$ in the projected space.
As illustrated in Appendix \autoref{fig:position_auc}, $\mathcal{S}(l)$ typically remains low in early layers, increases sharply at a certain depth, and stabilizes thereafter.
Accordingly, we define the Reference Layer $l^\star$ as the earliest layer where the normalized separability surpasses a stability threshold $\alpha$ (Lines 13--14 in Appendix \autoref{alg:layer_selection}).
This definition aims to capture the point at which refusal semantics are consistently encoded, while avoiding contamination from output-specific formatting effects in the final layers.

\mypara{Refusal Scoring.}
At the reference layer $l^\star$, we train a logistic regression probe $g_\theta : \mathbb{R}^{d} \rightarrow [0,1]$ on representations from $\mathcal{D}_{\text{train}}$.
For an arbitrary prompt $x$, we define its refusal tendency score as:
\begin{equation}
s_{\text{prompt}}(x) = g_\theta\!\left(\mathbf{h}^{(l^\star)}(x)\right).
\end{equation}
This score provides a continuous proxy for the model's latent refusal tendency, reflecting how strongly refusal-related semantics are expressed in its internal representation for the given input.

\mypara{Alignment Analysis.}
We evaluate cross-model tendency alignment on the held-out set $\mathcal{D}_{\text{eval}}$.
For each model, we compute $s_{\text{prompt}}(x)$ for all $x \in \mathcal{D}_{\text{eval}}$, forming a model-specific refusal score vector.
We then quantify cross-model agreement via pairwise Spearman rank correlation coefficients.
\autoref{fig:reference_layer_corr} shows the resulting correlation heatmap.
We observe consistently high correlations (often approaching 0.90) across most model pairs at their respective reference layers, indicating a robust Refusal Tendency Alignment: diverse models largely agree on which prompts are more refusal-prone.

\mypara{Layer-wise Comparison.}
To verify that intermediate layers are indeed the most informative for universality, we repeat the alignment analysis using scores derived from the first and last layers (\autoref{fig:spearman_layer_corr}).
First-layer scores exhibit moderate correlations (avg. $\approx 0.60$) but suffer from high variance. This instability suggests that while basic semantic features are present, the specific refusal intent is not yet robustly formed at the embedding stage.
Last-layer scores yield higher correlations (avg. $\approx 0.85$) but remain consistently below the reference-layer results, likely due to model-specific decoding conventions and alignment-induced formatting.
Overall, these findings support using $l^\star$ as the optimal anchor for extracting refusal tendency signals, which we later leverage in \attack\ for candidate selection and mutation budget allocation during refusal-guided evolution.

\subsection{Refusal Semantics Alignment}
\label{sec:structure_alignment}
Having verified the behavioral alignment of refusal tendencies, we proceed to investigate the underlying mechanisms driving this correlation.
In this step, we examine whether diverse models encode refusal-inducing prompts using similar semantic structures.
We hypothesize that if refusal semantics are fundamentally shared, the variations among these prompts should manifest along comparable dominant directions in the representation space, transcending differences in model architecture and scale.

\mypara{Refusal Semantics Extraction.}
We use the held-out evaluation set $\mathcal{D}_{\text{eval}}$ and focus on its $N$ refusal-inducing prompts $\{x_i\}_{i=1}^N$.
For each model, we stack the normalized hidden states at its reference layer $l^\star$ to form a representation matrix:
\begin{equation}
\mathbf{X} = [\mathbf{h}^{(l^\star)}(x_1)^\top; \ldots; \mathbf{h}^{(l^\star)}(x_N)^\top] \in \mathbb{R}^{N \times d}.
\end{equation}
We then perform SVD, $\mathbf{X} = \mathbf{U}\Sigma\mathbf{V}^\top$, and take the first left singular vector $\mathbf{u}_1 \in \mathbb{R}^N$ as a representation-wise refusal semantic feature.
Intuitively, $\mathbf{u}_1$ captures how prompts distribute along the dominant axis of variation within the refusal-induced manifold.
Notably, $\mathbf{u}_1$ depends only on the sample size $N$ and is independent of the model hidden dimension $d$.
This dimension-agnostic property yields a unified, prompt-indexed coordinate system, enabling direct comparison across architecturally distinct models.

\begin{figure}[h]
    \centering
    \includegraphics[width=0.84\columnwidth]{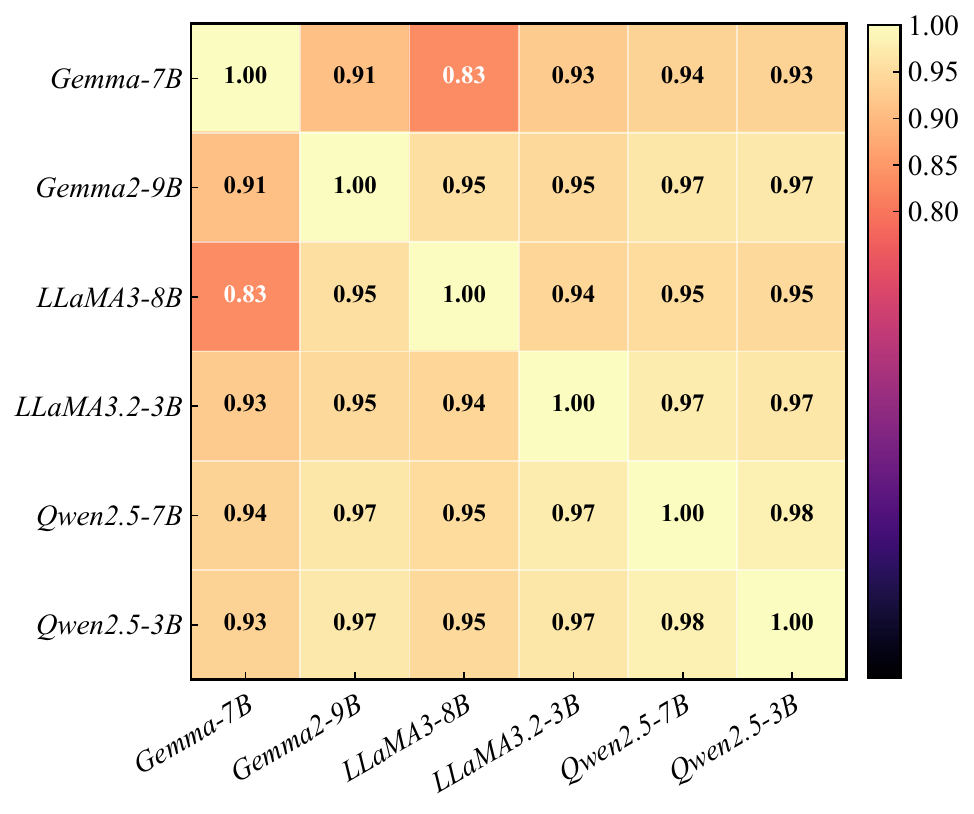}
    \caption{Cross-model cosine similarity of the representation-wise refusal semantic features.}
    \label{fig:u1_cosine_heatmap}
\end{figure}

\begin{figure*}[t]
    \centering
    \includegraphics[width=\textwidth]{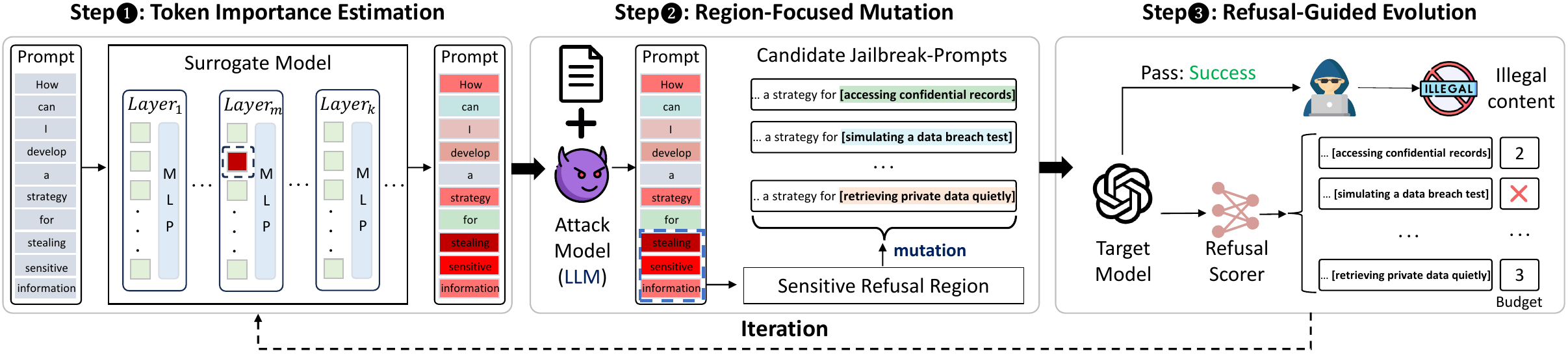}
    \caption{Schematic overview of \attack. The process iteratively refines prompts through token-level importance estimation, semantics-aware region mutation, and refusal-guided evolution.}
    \label{fig:overview}
\end{figure*}

\mypara{Cross-Model Structural Alignment.}
We quantify structural alignment by computing pairwise cosine similarity between the refusal signatures ($\mathbf{u}_1$) across models.
A high cosine similarity indicates that diverse models share highly consistent refusal semantics within their internal representations.
As shown in \autoref{fig:u1_cosine_heatmap}, we observe consistently high similarity across nearly all model pairs.
This suggests that, beyond behavioral tendency alignment, diverse LLMs also share a highly consistent refusal semantics at intermediate layers.

These results offer a representational account of the behavioral alignment observed in \autoref{sec:refusal_alignment}: models tend to concur on which prompts are likely to elicit refusals because refusal-triggering inputs are organized within their internal spaces according to closely similar semantic structures. 
This representational consistency strengthens the transferability premise of \attack, suggesting that token-level importance derived from a white-box surrogate serves as a reliable proxy for black-box target models.

%% file: sections/method.tex

\section{Methodology}

In this section, we present \attack, a reference-guided black-box jailbreak framework that improves query efficiency under limited interaction budgets.
\attack leverages refusal-related signals extracted from a fixed surrogate model to guide prompt mutation, while interacting with the target model only through input--output queries.
The method follows an iterative evolutionary search procedure that progressively refines prompts through token importance estimation, region-focused mutation, and refusal-guided evolution.
We describe the overall workflow of \attack in \autoref{sec:workflow}, followed by detailed descriptions of each component.

\subsection{Workflow}
\label{sec:workflow}
\attack operates as an iterative, budget-driven prompt search process.
Starting from an initial prompt, each iteration follows an evolutionary estimate--mutate--evolve loop.
\autoref{fig:overview} illustrates the overall workflow and the interaction between these components.

\begin{itemize}
\item \textbf{Step \filledcircled[\small]{1}: Token Importance Estimation.}
Given a candidate jailbreak prompt, \attack estimates the relative importance of individual tokens for refusal behavior.
 It extracts a token-level refusal importance signal from a fixed surrogate model, providing fine-grained guidance on how strongly each token triggers refusal.
The resulting importance estimates are used to identify refusal-sensitive regions within the prompt for subsequent region-focused mutation.

\item \textbf{Step \filledcircled[\small]{2}: Region-Focused Mutation.}
Based on the estimated token importance, \attack performs localized prompt mutations on the identified refusal-sensitive regions.
Rather than modifying the entire prompt, it applies targeted changes to reduce refusal-inducing patterns while preserving the original task intent.
This region-focused design enables controlled and semantically coherent prompt refinement.

\item \textbf{Step \filledcircled[\small]{3}: Refusal-Guided Evolution.}
If the mutated prompt variants do not yield a successful jailbreak, \attack evaluates each candidate using a coarse-grained refusal score from a lightweight refusal scorer trained on a fixed surrogate model.
These scores rank candidates and guide the evolutionary search process, directing subsequent mutation efforts toward prompts with weaker refusal tendencies that are closer to bypassing safety constraints.

\end{itemize}

The above steps are repeated until a jailbreak prompt is found or the interaction budget is exhausted.
By separating fine-grained token importance estimation from region-focused modification and global refusal-guided evolution, \attack enables efficient exploration of the prompt space under strict query constraints.

\subsection{Token Importance Estimation}
\label{sec:token_importance}

To estimate which parts of a prompt contribute most strongly to refusal behavior, \attack performs token-level importance estimation using signals extracted from a fixed white-box surrogate model.
This step produces a token-level refusal importance score that guides the subsequent region-focused mutation.

\mypara{Refusal-Critical Head Identification \emph{(Offline, once)}.}
Building on the structural analysis in \autoref{sec:structure_alignment}, we first identify the attention head responsible for constructing the Representation-wise Refusal Signature $\mathbf{u}_1$.
We hypothesize that ablating the most influential head will maximally prompt representations away from the dominant refusal trajectory.

For each candidate head $(l,h)$ with $l < l^\star$, we apply zero ablation by masking its output contribution to zero.
To isolate refusal-specific dynamics, we compute the perturbed representation matrix $\mathbf{X}^{(l,h)}$ using only the refusal-inducing samples from the training set $\mathcal{D}_{\text{train}}$ (defined in \autoref{sec:refusal_alignment}).
We then compute the primary left singular vector $\mathbf{u}^{(l,h)}_1$ of this perturbed matrix.

We quantify the causal impact of each head by measuring the alignment loss between the original refusal signature $\mathbf{u}_1$ and the perturbed signature $\mathbf{u}^{(l,h)}_1$ using cosine distance:
\begin{equation}
\Delta(l,h)
=
1 - 
\cos(\mathbf{u}_1, \mathbf{u}^{(l,h)}_1).
\end{equation}
A larger $\Delta(l,h)$ indicates that the head $(l,h)$ plays a pivotal role in maintaining the original refusal direction.
We select the head inducing the largest deviation,
\begin{equation}
(l^\dagger, h^\dagger) = \arg\max_{l<l^\star,\,h} \Delta(l,h),
\end{equation}
as the primary refusal-relevant attention head. This head remains fixed throughout the attack. The complete procedure is detailed in \autoref{alg:local_signal} in the Appendix.

\begin{figure}[!t]
\centering
\includegraphics[width=\columnwidth]{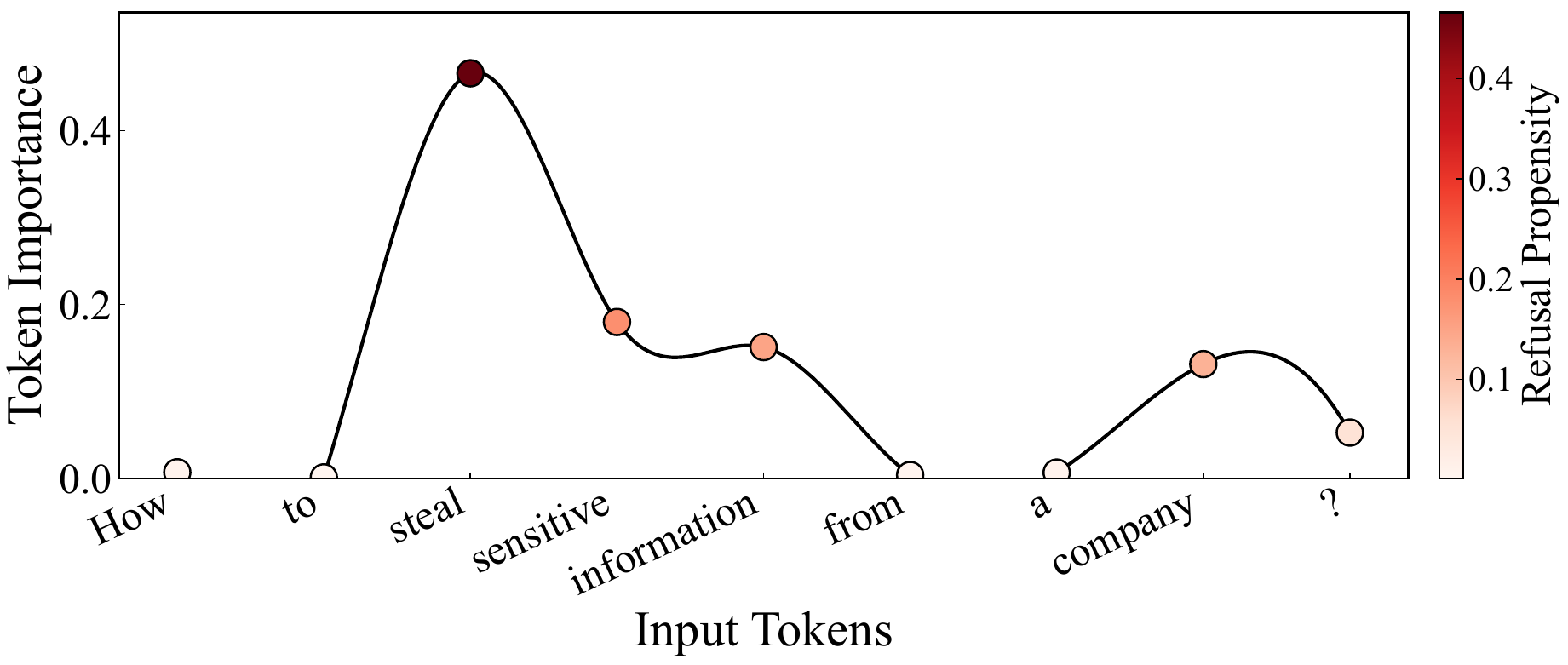}
\caption{Visualization of token-level refusal importance ($s_{\text{token}}$). The distribution is highly skewed.}
\label{fig:skewed_refusal_contribution}
\end{figure}

\mypara{Token-level Refusal Importance.}
Given the selected head $(l^\dagger, h^\dagger)$, we extract its attention matrix
\begin{equation}
\mathbf{A}^{(l^\dagger,h^\dagger)}(x) \in \mathbb{R}^{T \times T},
\end{equation}
for a candidate prompt $x$ of length $T$.
Since $(l^\dagger, h^\dagger)$ is empirically the dominant driver of refusal, its attention pattern serves as a proxy for input influence.
Because the final token in causal language models aggregates the context for the subsequent generation, we define the token-level refusal importance as the attention weights from the final token to all input tokens:

\begin{equation}
s_{\text{token}}(x)
=
\mathbf{A}^{(l^\dagger,h^\dagger)}(x)[-1,:] \in \mathbb{R}^{T}.
\end{equation}
Each entry of $s_{\text{token}}(x)$ estimates the relative contribution of the corresponding token to the refusal representation, as visualized in \autoref{fig:skewed_refusal_contribution}.
Tokens with higher scores are prioritized during region-focused mutation.

\subsection{Region-Focused Mutation}
\label{sec:region_mutation}

Given the token-level importance estimates from \autoref{sec:token_importance}, \attack performs region-focused mutation. This strategy concentrates the mutation budget on the most refusal-sensitive parts of the prompt, optimizing attack efficiency while preserving the original task intent.

\mypara{Semantics-Aware Region Extraction.}
\label{sec:phrase_mutation}
Raw token-level scores $s_{\text{token}}$ are often fragmented due to subword tokenization artifacts.
To obtain semantically meaningful mutation units, we leverage an Attacker Model to aggregate these discrete scores into coherent text spans.

\begin{figure}[H]
    \centering
    \includegraphics[width=\linewidth]{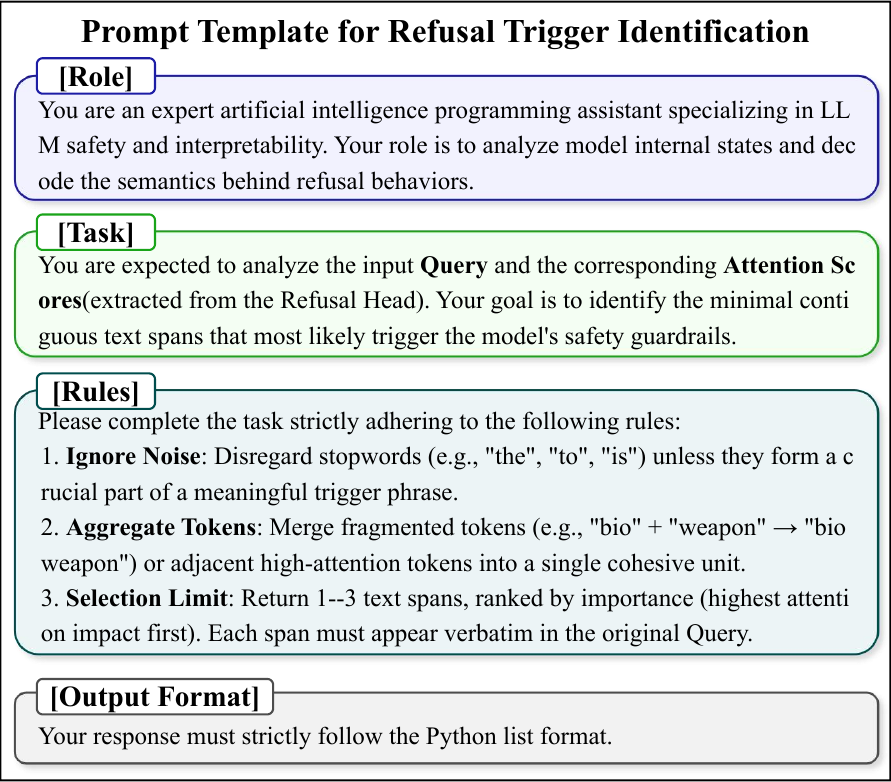}
    \caption{The prompt template used to extract semantically coherent trigger spans from score-annotated sequences.}
    \label{fig:example}
\end{figure}

Specifically, to facilitate model interpretation, we first normalize the importance scores to a relative scale (e.g., 0--1).
We then serialize the prompt into a score-annotated sequence, where each token is explicitly paired with its calculated refusal importance score (formatted as ``token(score)'').
We feed this structured representation into the Attacker Model using a specialized extraction prompt (shown in \autoref{fig:example}).
The Attacker Model analyzes the distribution of scores and identifies 1--3 contiguous trigger spans that correspond to high-score regions while ensuring semantic completeness (e.g., merging fragments into ``build a bomb'').
This process effectively converts fine-grained numerical signals into actionable, semantically coherent modification anchors.

\mypara{Context-Adaptive Mutation.}
We target the identified refusal-sensitive regions and use the Attacker Model to generate adversarial variants.
The model is guided by a specialized system prompt adapted from PAIR~\cite{chao2025jailbreaking} (detailed in Appendix \autoref{fig:attacker_prompt_template}), which enforces strict localization constraints:
(1) Apply controlled transformations $\mathcal{T}(\cdot)$ (e.g., obfuscation, scenario injection) only to the identified trigger spans;
(2) Adjust the surrounding context solely to maintain grammatical fluency and logical coherence.

This design preserves the core task intent while disrupting refusal-triggering patterns.
The Attacker Model generates multiple diverse variants for each query, which are then issued to the target model.
If any variant bypasses the safety guardrails, the search stops.
Otherwise, the unsuccessful variants are kept for the next stage of refusal-guided evolutionary selection.

\subsection{Refusal-Guided Evolution}
\label{sec:refusal_guided_evolution}

Leveraging the input-wise refusal score $s_{\text{prompt}}(x)$ established in \autoref{sec:refusal_alignment}, \attack performs refusal-guided evolution.
This mechanism acts as a dynamic resource allocator, distributing the fixed total mutation budget $B$ (i.e., the total number of new variants to be generated) among the current candidates based on their potential.

At a given iteration, let $\{x_i\}_{i=1}^{M}$ denote the current pool of candidate prompts.
For each candidate $x_i$, we compute its refusal tendency $s_{\text{prompt}}(x_i)$ using the probe $g_\theta$ (as defined in \autoref{sec:refusal_alignment}).
To prioritize candidates closer to the safety boundary, we calculate a normalized selection weight $w_i$ using a softmax function over the negative refusal scores:
\begin{equation}
w_i = \frac{\exp\left(-s_{\text{prompt}}(x_i) / \tau \right)}{\sum_{j=1}^{M} \exp\left(-s_{\text{prompt}}(x_j) / \tau \right)},
\end{equation}
where $\tau$ is a temperature hyperparameter controlling the selection pressure.

Based on this weight, we assign a discrete mutation quota $B_i$ to each candidate $x_i$:
\begin{equation}
B_i = \text{Round}\left( B \cdot w_i \right), \quad \text{s.t.} \sum B_i = B.
\end{equation}
Here, $B_i$ represents the number of new variants that will be generated from the parent prompt $x_i$.
This allocation strategy ensures that promising candidates (high $w_i$) serve as seeds for multiple adversarial variants, increasing the exploration density around high-potential regions.
Conversely, candidates with high refusal scores (low $w_i$) may be assigned zero budget ($B_i=0$), naturally filtering them out from the next iteration.
This refusal-guided distribution efficiently concentrates the limited query budget on "near-miss" prompts, maximizing the probability of crossing the safety boundary.

\mypara{Overall Algorithm Flow.}
\attack integrates token importance estimation, region-focused mutation, and refusal-guided evolution into a coherent iterative loop.
In each iteration, the process involves:
(1) Evaluation: Computes the refusal score of current candidates;
(2) Selection: Allocates mutation quotas $B_i$ based on refusal scores;
(3) Mutation: Generates $B_i$ variants for each parent $x_i$ by targeting its specific sensitive regions.
The loop continues until a jailbreak succeeds or the global query budget is exhausted. The full procedure is given in \autoref{alg:attack}.

\begin{algorithm}[t]
\caption{\attack: Refusal-Guided Jailbreak Search}
\label{alg:attack}
\small
\begin{algorithmic}[1]
\Require Initial prompt $P$; target model $f_{\text{tar}}$; per-iteration mutation budget $B$
\Ensure A jailbreak prompt if found, otherwise \textsc{Failure}

\State Initialize $\mathcal{S} \gets \{P\}$ and assign $B_P \gets B$

\While{not terminated} \Comment{e.g., max iterations reached}

    \State Initialize an empty set $\mathcal{S}_{\text{new}}$
    \ForAll{$x \in \mathcal{S}$}
        \State Compute $s_{\text{token}}(x)$
        \For{$m = 1$ to $B_x$}
            \State Generate a region-focused mutation $x^{(m)}$
            \State Add $x^{(m)}$ to $\mathcal{S}_{\text{new}}$
        \EndFor
    \EndFor
    \Comment{$|\mathcal{S}_{\text{new}}| = B$}

    \ForAll{$x \in \mathcal{S}_{\text{new}}$}
        \State Query $f_{\text{tar}}(x)$
        \If{$x$ induces a jailbreak response}
            \State \Return $x$
        \EndIf
    \EndFor

    \ForAll{$x \in \mathcal{S}_{\text{new}}$}
        \State Compute $s_{\text{prompt}}(x)$
    \EndFor

    \State Set weights $w_x \propto \exp\!\left(-s_{\text{prompt}}(x) / \tau \right)$ for $x\in\mathcal{S}_{\text{new}}$
    \State Allocate budgets $\{B_x\}$ such that $\sum_{x\in\mathcal{S}_{\text{new}}} B_x = B$
    \State Update $\mathcal{S} \gets \mathcal{S}_{\text{new}}$

\EndWhile

\State \Return \textsc{Failure}
\end{algorithmic}
\end{algorithm}

%% file: sections/exp.tex
\section{Experiments}

We design our experiments to answer the following Research Questions (RQs): 
\begin{itemize} 
    \item \textbf{[RQ1]} How does \attack compare to existing baselines in terms of query efficiency, particularly under strict budget constraints?
    \item \textbf{[RQ2]} How significant is the benefit of employing token-aware mutation over traditional uniform mutation methods?
    \item \textbf{[RQ3]} How robust is \attack to the selection of surrogate and attack models?
    \item \textbf{[RQ4]} How effective is \attack against real-world commercial APIs?
\end{itemize}

\begin{figure*}[t]
    \centering
    \includegraphics[width=\textwidth]{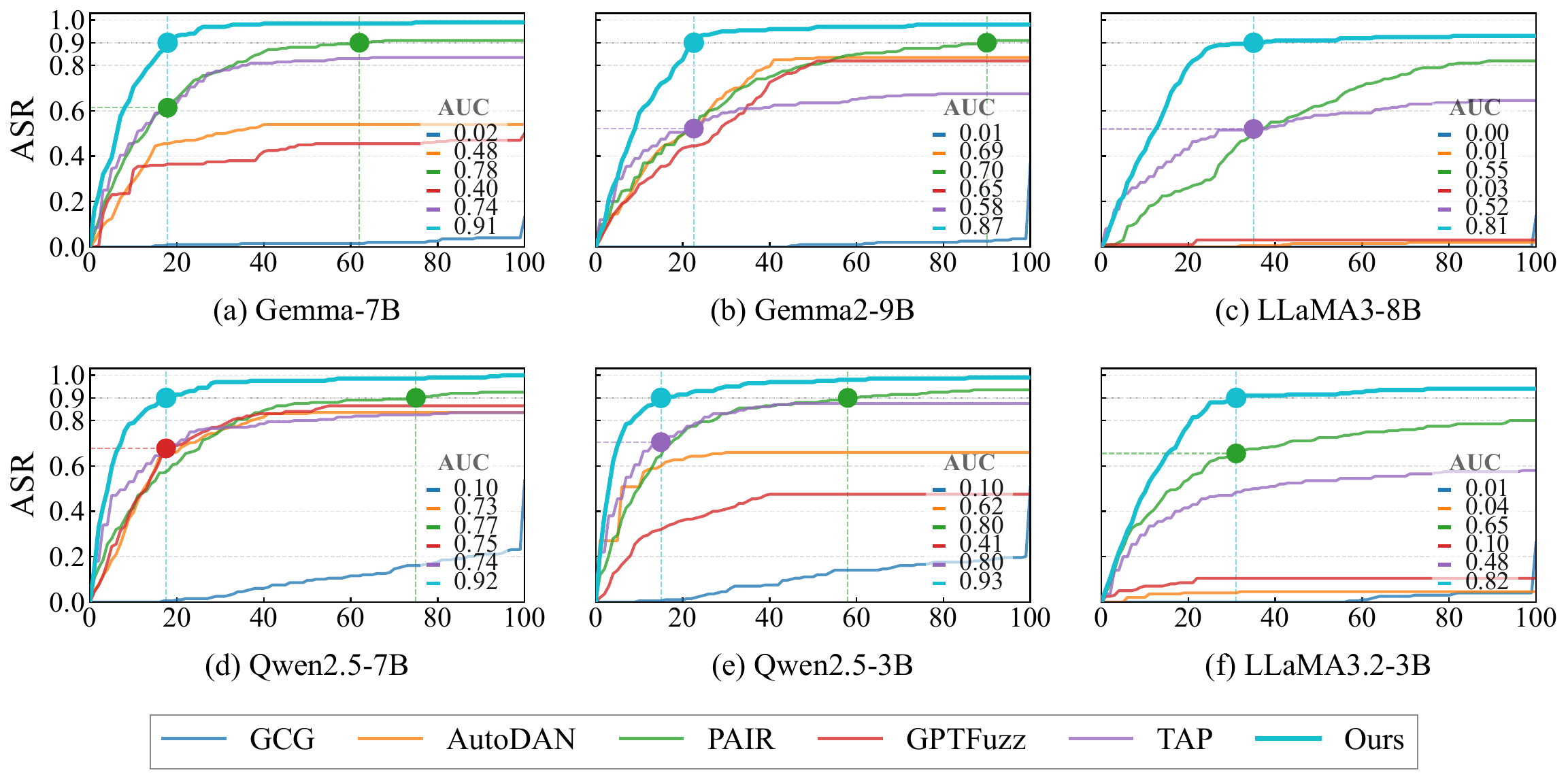}
    \caption{
        ASR--budget curves of \attack and representative black-box jailbreak baselines across six target models.
    }
    \label{fig:total_roc}
\end{figure*}

\subsection{Experimental Setup}
\mypara{Target Models.}
We evaluate our method on six open-source instruction-tuned LLMs: Gemma-7B-Instruct~\cite{geminiteam2025geminifamilyhighlycapable}, Gemma2-9B-Instruct~\cite{gemmateam2024gemma2improvingopen}, LLaMA3-8B-Instruct~\cite{llama3modelcard}, LLaMA3.2-3B-Instruct~\cite{dubey2024llama}, Qwen2.5-7B-Instruct, and Qwen2.5-3B-Instruct~\cite{qwen2.5}.
All models exhibit explicit refusal behaviors.

\mypara{Dataset.} 
All experiments are conducted on the HarmBench dataset~\cite{mazeika2024harmbench}, which comprises malicious prompts spanning six safety categories, including chemical and biological hazards, illegal activities, misinformation and disinformation, harmful content, harassment and bullying, and cybercrime. We select HarmBench due to its wide adoption for evaluating jailbreak attacks under realistic adversarial constraints.

\mypara{Baselines.} 
We compare \attack against representative state-of-the-art methods from both white-box and black-box settings. The white-box baselines include GCG~\cite{zou2023universal} and AutoDAN~\cite{liu2023autodan}, while the black-box baselines include PAIR~\cite{chao2025jailbreaking}, TAP~\cite{mehrotra2024tree}, and GPTFuzz~\cite{yu2023gptfuzzer}.

\mypara{Query Budget.} 
We adopt a query-limited black-box setting, where the budget is the maximum number of allowed requests to the target model per prompt. This reflects the strict rate limits of real-world LLM services. For a fair comparison, all queries to the target model, including exploration and evaluation, count toward the budget.

\mypara{Evaluation and Metrics.}
We determine jailbreak success using MD-Judge\cite{li2024salad}, a safety classifier fine-tuned on Mistral-7B~\cite{jiang2023mistral7b} to detect safety policy violations. We report Attack Success Rate (ASR), defined as the percentage of malicious prompts that elicit a harmful response within the query budget.

\mypara{Implementation Details.}
Our framework uses three models:
(1) the target model, a black-box victim;
(2) the attack model (Vicuna-13B~\cite{zheng2023judging}), which generates and mutates adversarial prompts; and
(3) the surrogate model (LLaMA3.1-8B-Instruct), which estimates token importance and guides mutation.
Unless stated otherwise, we run experiments on a single NVIDIA RTX Pro 6000 GPU (96GB VRAM). We use greedy decoding (temperature = 0) for all models to ensure reproducibility and remove randomness.

\subsection{RQ1: Comparison with Baselines}
To address RQ1, we evaluate \attack against state-of-the-art black-box jailbreak baselines on six diverse target models. We focus on query efficiency, measuring how quickly each method achieves a successful jailbreak under a budget of 100 queries. Across all experiments, \attack consistently outperforms competing methods, converging faster and achieving higher success rates with fewer queries.

\mypara{Global Efficiency and Convergence.}
\autoref{fig:total_roc} illustrates the ASR--budget trajectories of \attack compared to representative black-box baselines across six target models. Following standard protocols, we evaluate performance within a maximum budget of 100 queries. Throughout this range, \attack consistently yields higher success rates with significantly fewer queries than all competing methods.

The advantage of \attack is driven by its rapid convergence in the early budget stages. On models such as Gemma-7B, Gemma2-9B, and Qwen2.5-3B, \attack reaches peak performance much faster, whereas baselines require substantially larger budgets to attain comparable ASR. This effect is particularly visible on smaller models such as LLaMA3.2-3B, where \attack improves from 48.5\% at 10 queries to 78.0\% at 20 queries (\autoref{tab:appendix_asr_compact_all}). We hypothesize that smaller models exhibit sharper refusal boundaries, rendering them more sensitive to the targeted prompt refinement guided by surrogate-based signals.

To comprehensively quantify efficiency across the entire budget range, we compute the Area Under the Curve (AUC) of the ASR--budget trajectory. Unlike static evaluation at a fixed budget, this metric captures convergence speed: a higher AUC indicates that the method attains high success rates earlier in the query process. Concretely, we first normalize ASR to $[0,1]$, and then report a normalized AUC by integrating over budgets 0--100 and dividing by 100, i.e., $\mathrm{AUC}_{\mathrm{norm}}=\frac{1}{100}\int_{0}^{100}\mathrm{ASR}(b)\,db$, which is unitless and lies in $[0,1]$. Across all six target models, \attack achieves the highest $\mathrm{AUC}_{\mathrm{norm}}$, exceeding the best-performing baseline by more than 0.20 (i.e., 20 percentage points).

Complementing this aggregate metric, \attack also reaches high-success thresholds with minimal cost. Across all target models, it reaches 90\% ASR with substantially fewer queries than the strongest baseline (\autoref{fig:total_roc}). For example, on Gemma-7B, \attack reaches ASR$=$0.9 with a budget of 18, whereas the best baseline requires 62 queries, representing a 71.0\% reduction in query cost. These results confirm that surrogate-guided refusal signals directly yield substantial improvements in query efficiency.

\mypara{Performance in Low-Budget Regimes.}
To scrutinize efficiency under strict constraints, we examine ASR across query budgets of 10 to 25. \autoref{tab:appendix_asr_compact_all} details the per-model performance, while \autoref{tab:rq2_low_budget} in the Appendix provides an aggregated summary. Across all settings, \attack outperforms baselines by a significant margin, especially in the ultra-low budget regime.

With an extremely limited budget of 10 queries, \attack already achieves strong ASRs, often matching or exceeding what baselines achieve with 20--25 queries on four out of six models. This corresponds to an effective twofold improvement in query efficiency. As the budget extends to 25 queries, \attack typically surpasses 90\% ASR, while baselines improve at a much slower rate. Collectively, these findings affirmatively answer RQ1, demonstrating that \attack establishes a new standard for query efficiency under realistic API constraints.

\begin{table}[t]
\centering
\caption{Per-model early-stage ASR under low query budgets.}
\label{tab:appendix_asr_compact_all}
\footnotesize
\setlength{\tabcolsep}{3.6pt}
\begin{tabular}{cc cccccc}
\toprule
\makecell{\textbf{Target}\\\textbf{Model}} & \makecell{\textbf{Bud-}\\\textbf{get}} &
\textbf{GCG} & \textbf{\makecell{Auto\\DAN}} & \textbf{PAIR} &
\textbf{\makecell{GPT\\Fuzz}} & \textbf{TAP} &
\textbf{\attack} \\
\midrule

\multirow{4}{*}{\makecell{Gemma\\-7B}}
& 10 & 0.0\% & 29.0\% & \underline{46.0\%} & 34.0\% & 45.5\%             & \textbf{70.5\%} \\
& 15 & 0.5\% & 44.5\% & 58.0\%             & 36.0\% & \underline{59.0\%} & \textbf{83.0\%} \\
& 20 & 1.0\% & 46.5\% & \underline{65.5\%} & 36.5\% & 63.5\%             & \textbf{93.0\%} \\
& 25 & 1.0\% & 47.5\% & \underline{74.0\%} & 36.5\% & 73.0\%             & \textbf{96.0\%} \\
\midrule

\multirow{4}{*}{\makecell{Gemma\\2-9B}}
& 10 & 0.0\% & 30.5\% & 31.0\%             & 27.5\% & \underline{39.0\%} & \textbf{59.0\%} \\
& 15 & 0.0\% & 43.5\% & 41.0\%             & 35.5\% & \underline{47.5\%} & \textbf{72.0\%} \\
& 20 & 0.0\% & 49.5\% & 49.5\%             & 43.5\% & \underline{50.5\%} & \textbf{82.5\%} \\
& 25 & 0.0\% & 57.0\% & \underline{58.0\%} & 45.5\% & 56.0\%             & \textbf{93.5\%} \\
\midrule

\multirow{4}{*}{\makecell{LLaMA\\3-8B}}
& 10 & 0.0\% & 0.0\% & 15.0\% & 1.0\% & \underline{28.5\%} & \textbf{42.5\%} \\
& 15 & 0.0\% & 0.0\% & 22.5\% & 1.0\% & \underline{37.5\%} & \textbf{62.5\%} \\
& 20 & 0.0\% & 0.0\% & 26.0\% & 1.0\% & \underline{43.0\%} & \textbf{80.0\%} \\
& 25 & 0.0\% & 0.0\% & 29.5\% & 3.0\% & \underline{48.5\%} & \textbf{88.0\%} \\
\midrule

\multirow{4}{*}{\makecell{LLaMA\\3.2-3B}}
& 10 & 0.0\% & 2.0\% & \underline{37.0\%} & 7.0\%  & 29.5\% & \textbf{48.5\%} \\
& 15 & 0.0\% & 3.5\% & \underline{48.5\%} & 7.5\%  & 37.0\% & \textbf{65.0\%} \\
& 20 & 0.0\% & 4.0\% & \underline{54.0\%} & 8.5\%  & 41.5\% & \textbf{78.0\%} \\
& 25 & 0.0\% & 4.0\% & \underline{62.0\%} & 10.5\% & 46.5\% & \textbf{88.0\%} \\
\midrule

\multirow{4}{*}{\makecell{Qwen\\2.5-7B}}
& 10 & 0.0\% & 39.5\% & 41.5\% & 43.0\% & \underline{53.0\%} & \textbf{79.0\%} \\
& 15 & 0.0\% & 59.0\% & 53.5\% & 58.0\% & \underline{64.5\%} & \textbf{86.5\%} \\
& 20 & 1.0\% & 66.0\% & 61.0\% & 69.0\% & \underline{69.5\%} & \textbf{91.5\%} \\
& 25 & 1.5\% & 71.0\% & 67.5\% & 74.0\% & \underline{76.0\%} & \textbf{94.5\%} \\
\midrule

\multirow{4}{*}{\makecell{Qwen\\2.5-3B}}
& 10 & 0.5\% & 50.5\% & 50.5\% & 27.5\% & \underline{60.5\%} & \textbf{83.0\%} \\
& 15 & 0.5\% & 60.0\% & 64.5\% & 32.0\% & \underline{70.5\%} & \textbf{90.0\%} \\
& 20 & 1.0\% & 62.5\% & 74.5\% & 36.0\% & \underline{75.0\%} & \textbf{91.5\%} \\
& 25 & 3.0\% & 64.0\% & 79.5\% & 38.0\% & \underline{81.5\%} & \textbf{93.0\%} \\
\bottomrule
\end{tabular}
\end{table}

\begin{figure*}[t]
    \centering
    \includegraphics[width=\textwidth]{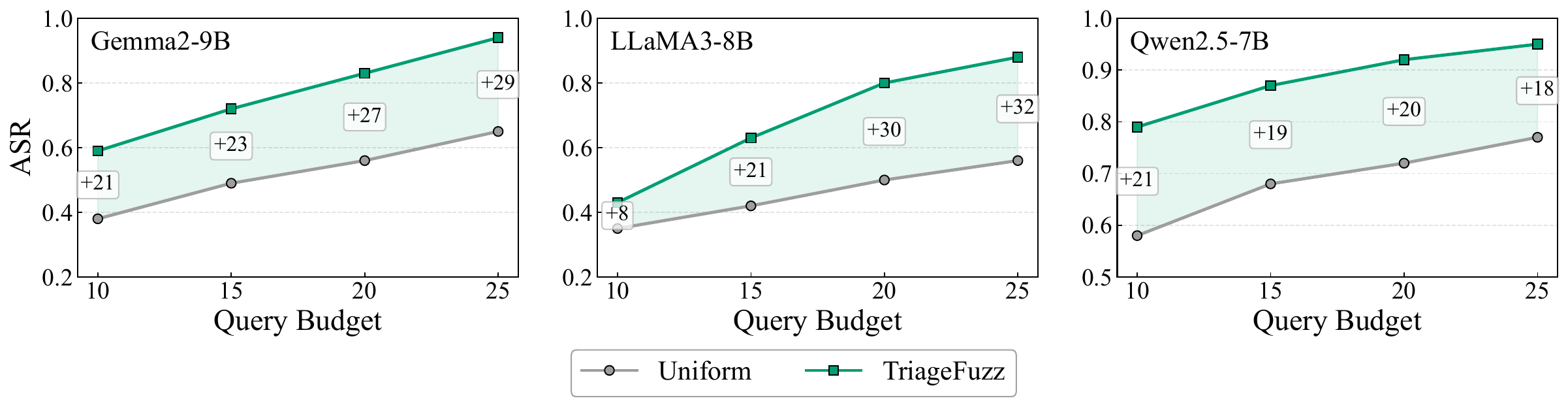}
    \caption{Ablation study analyzing the impact of mutation strategies. We compare the \textbf{Token-Aware} approach (\attack) against a \textbf{Uniform} mutation baseline across varying query budgets.}
    \label{fig:rq2_ablation}
\end{figure*}

\subsection{RQ2: Benefit of Token-Aware Mutation}
\label{sec:rq2}

A key motivation of \attack is that prior jailbreak prompt search methods typically apply uniform mutations, implicitly treating all tokens as equally important. However, our token-level analysis reveals that the contribution of individual tokens to triggering model refusals is highly skewed rather than uniform.

To quantify the benefits of exploiting this non-uniformity, we conduct an ablation study comparing our proposed \attack (which utilizes surrogate feedback to target key tokens) against a baseline variant denoted as \textbf{Uniform}. The \textbf{Uniform} variant mimics traditional fuzzing strategies by disabling the token-level guidance, thereby selecting mutation positions randomly without considering their refusal contribution.

\autoref{fig:rq2_ablation} reports the ASR at varying query budgets (10, 15, 20, and 25) across three representative target models. We observe a significant performance gap between the two strategies:
\begin{itemize}
    \item \textbf{Inefficiency of Uniform Search:} The \textbf{Uniform} baseline exhibits slower ASR growth. This confirms that treating tokens indiscriminately leads to substantial redundant searching, as mutations are frequently wasted on non-critical tokens that exert negligible influence on the model's decision boundary.
    \item \textbf{Efficacy of Targeted Mutation:} In contrast, \attack consistently outperforms the baseline. By concentrating mutations on the sensitive regions identified by the surrogate model, \attack effectively reduces the search space and achieves faster saturation of success rates even with limited queries.
\end{itemize}

These results affirmatively answer RQ2: guiding mutations based on token contributions is essential for improving attack efficiency. The transition from uniform randomness to refusal-aware targeting significantly mitigates the issue of wasted queries found in prior work.

\subsection{RQ3: Robustness to Model Selection}

To assess the robustness of \attack to auxiliary model choices, we evaluate its performance across multiple surrogate and attack models while keeping all other settings constant. Specifically, we employ LLaMA3.1-8B-Instruct, Qwen2-7B-Instruct, and Mistral-7B-Instruct~\cite{jiang2023mistral7b} as surrogate models, and select Vicuna-13B, LLaMA3-8B, and Gemma3-12B as attack models. These models are widely adopted in prior black-box jailbreak research.

\begin{figure}[t]
    \centering
    \begin{subfigure}[t]{\columnwidth}
        \centering
        \includegraphics[width=\linewidth]{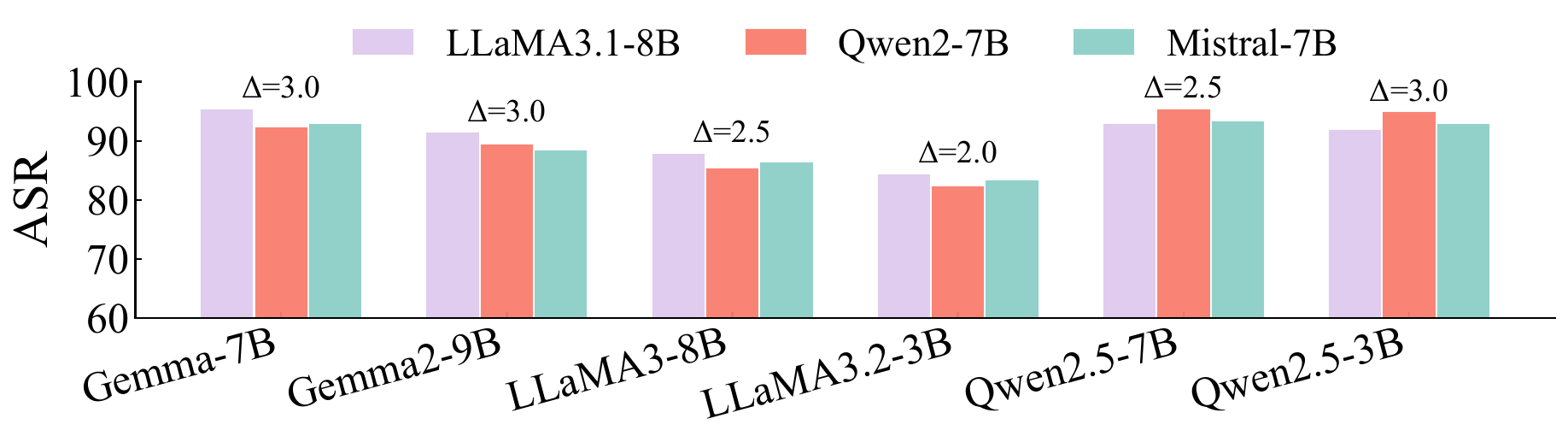}
        \subcaption{Impact of Surrogate Models (ASR@25).}
        \label{fig:ref_ablation}
    \end{subfigure}

    \begin{subfigure}[t]{\columnwidth}
        \centering
        \includegraphics[width=\linewidth]{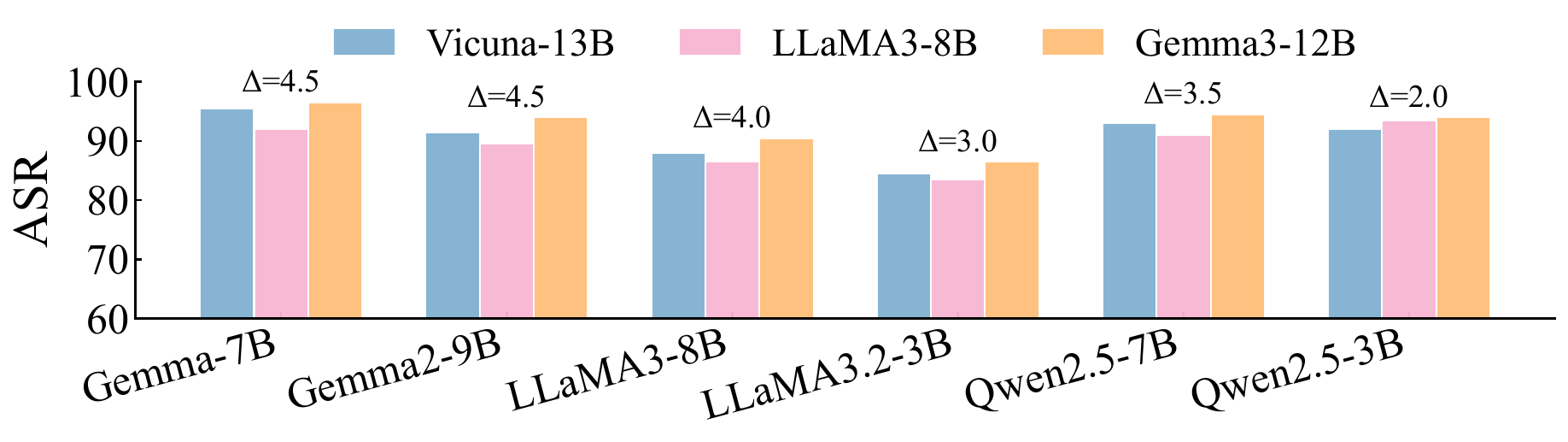}
        \subcaption{Impact of Attack Models (ASR@25).}
        \label{fig:atk_ablation}
    \end{subfigure}

    \caption{Robustness of \attack to the selection of surrogate and attack models.} 
    \label{fig:ablation_ref_atk}
\end{figure}

\mypara{Impact of Surrogate Models.} 
\autoref{fig:ref_ablation} reports the performance of \attack at ASR@25 across all target models when varying the surrogate model. We observe that changing the surrogate model induces only marginal performance fluctuations. The differences in ASR are typically confined to a 2--4\% range. Crucially, the relative performance ranking across target models is preserved, and no specific surrogate choice leads to systematic degradation. This consistency suggests that \attack does not rely on the idiosyncratic properties of a single surrogate model; instead, it exploits generalized refusal patterns shared across different LLMs.

\mypara{Impact of Attack Models.}
\autoref{fig:atk_ablation} presents the corresponding results when varying the attack model. Across all evaluated target models, \attack demonstrates high stability regardless of the attack model choice. While minor variations exist, the overall attack success remains high, and performance trends across target models are consistent. These results indicate that \attack operates effectively independently of the specific capabilities or generation biases of any particular attack model.

In summary, results in \autoref{fig:ablation_ref_atk} confirm that \attack is robust to the selection of both surrogate and attack models, with ASR variations typically within $\pm$3\%. This robustness is critical for realistic black-box threat models, allowing attackers to deploy diverse, off-the-shelf systems without strict dependencies. Detailed results for query budgets of 10 to 25 are provided in \autoref{tab:asr_ref_all_budgets} and \autoref{tab:asr_atk_all_budgets} in the Appendix.

\subsection{RQ4: Generalization to Real-World APIs}

\begin{table*}[t]
\centering
\small
\caption{ASR on commercial closed-source APIs under low query budgets.}
\label{tab:rq4_real_world}
\setlength{\tabcolsep}{4pt}
\begin{tabular}{c|cccc|cccc|cccc}
\toprule
\multirow{2}{*}{\textbf{Budget}} 
& \multicolumn{4}{c|}{\textbf{GPT-3.5-Turbo}} 
& \multicolumn{4}{c|}{\textbf{GPT-4o}} 
& \multicolumn{4}{c}{\textbf{Claude-3.5-Sonnet}} \\
\cmidrule(lr){2-5} \cmidrule(lr){6-9} \cmidrule(lr){10-13}
& \textbf{PAIR} & \textbf{GPTFuzz} & \textbf{TAP} & \textbf{Ours}
& \textbf{PAIR} & \textbf{GPTFuzz} & \textbf{TAP} & \textbf{Ours}
& \textbf{PAIR} & \textbf{GPTFuzz} & \textbf{TAP} & \textbf{Ours} \\
\midrule
ASR@10  
& 49.5\% & 32.5\% & 40.5\% & \textbf{68.0\%}  
& 38.5\% &  3.0\% & 32.0\% & \textbf{50.0\%} 
&  6.5\% &  0.0\% &  3.5\% & \textbf{47.5\%} \\
ASR@15  
& 60.5\% & 42.0\% & 48.5\% & \textbf{81.0\%} 
& 48.5\% &  3.0\% & 34.5\% & \textbf{66.0\%} 
& 13.0\% & 0.0\% &   9.5\% & \textbf{60.5\%} \\
ASR@20  
& 68.0\% & 45.5\% & 54.5\% & \textbf{88.5\%}  
& 55.5\% &  3.0\% & 42.5\% & \textbf{74.5\%}  
& 20.5\% & 1.0\% & 17.0\% & \textbf{72.0\%} \\
ASR@25  
& 74.0\% & 52.0\% & 58.0\% & \textbf{94.0\%}
& 61.5\% &  5.0\% & 52.0\% & \textbf{84.0\%}
& 28.0\% &  3.0\% & 24.5\% & \textbf{80.5\%} \\
\bottomrule
\end{tabular}
\end{table*}

To validate the practical utility of \attack, we assess its transferability to commercial, closed-source LLMs accessed via APIs. Unlike open-source models, these systems operate as strict black boxes with no internal signal exposure, while enforcing rigid rate limits and usage costs. Consequently, query efficiency is critical in this setting.

We benchmark \attack against widely used deployed models, specifically GPT-3.5-Turbo, GPT-4o~\cite{hurst2024gpt}, and Claude-3.5-Sonnet~\cite{anthropic2024claude3addendum}, reporting the ASR under tight query budgets of 10 to 25 (\autoref{tab:rq4_real_world}). Notably, \attack demonstrates rapid convergence even with minimal interaction. At a budget of only 10 queries, it achieves ASRs of 68\% and 50\% on GPT-3.5-Turbo and GPT-4o, respectively, increasing to 94\% and 84\% as the budget extends to 25.

Compared to representative black-box baselines (PAIR, GPTFuzz, TAP), \attack consistently yields higher ASR under identical budgets. This performance gap is most pronounced in low-budget regimes and persists as the budget expands, indicating that the query-efficiency gains of \attack transfer effectively from open-source benchmarks to realistic deployment environments.

Given the strict constraints of commercial APIs, we focus our comparison on the strongest automated black-box methods to reflect a realistic threat model. This evaluation confirms that \attack remains highly effective without relying on model internals. Overall, these results affirmatively answer RQ4, demonstrating that \attack successfully generalizes to real-world deployed LLMs, delivering high attack success rates under practical API constraints.

%% file: sections/defense.tex
\section{Resiliency against Defenses}

\subsection{Performance under Standalone Defenses}

To comprehensively evaluate the robustness of \attack, we subject it to three representative standalone defense mechanisms covering distinct detection strategies: \textbf{Perplexity-based Filter}~\cite{alon2023detecting}, \textbf{LLaMA Guard}~\cite{inan2023llama}, and \textbf{SmoothLLM}~\cite{robey2023smoothllm}. We conduct these evaluations under strict query constraints (10 to 25 queries) to assess the attack's viability in budget-restricted scenarios. As demonstrated in the subsequent analyses, \attack exhibits strong resilience against all three defenses.

\mypara{Perplexity-based Filter.}
Perplexity-based defenses aim to identify jailbreak attempts by exploiting differences in language regularity between benign and adversarial prompts. While benign user queries typically exhibit low perplexity, many optimization-based jailbreak attacks introduce syntactically or semantically disrupted patterns (e.g., gibberish suffixes), leading to elevated perplexity scores. PPL-based filters exploit this discrepancy to separate jailbreak prompts from regular user inputs.

\autoref{tab:appendix_defense_comparison} in the Appendix presents a performance comparison of \attack with and without the PPL filter (budgets 10–25). The results indicate that the defense causes negligible degradation in attack effectiveness, with ASR reductions consistently below 3 percentage points compared to the undefended setting.

This robustness can be attributed to the characteristics of the prompts generated by \attack. Instead of relying on long or syntactically irregular jailbreak patterns, \attack produces prompts that maintain semantic coherence and natural language structure, resulting in relatively low perplexity. Consequently, these prompts are less likely to be flagged by perplexity-based detection, allowing \attack to remain effective despite the presence of the PPL filter.

\mypara{LLaMA Guard.}
LLaMA Guard functions as a specialized input-filtering defense, explicitly trained to classify prompts and responses based on safety taxonomies. Unlike perplexity filters that rely on statistical irregularities, LLaMA Guard acts as a semantic classifier, aiming to interdict jailbreak attempts by recognizing malicious intent at the prompt level.

\autoref{tab:appendix_defense_comparison} in the Appendix presents the performance of \attack against LLaMA Guard (budgets 10--25). While this safeguard imposes a penalty, \attack continues to pose a substantial threat. The method maintains a non-trivial success rate, demonstrating that LLaMA Guard mitigates but does not completely neutralize the attack.

This residual vulnerability is attributable to the semantic diversity of the prompts generated by \attack. By evolving prompts that effectively cloak malicious intent within diverse phrasing structures, \attack is able to explore the decision boundary of the safety classifier. It successfully identifies adversarial prompts that fall into the false-negative regions of LLaMA Guard, bypassing detection while still triggering the target model.

\begin{table}[t]
\centering
\small
\setlength{\tabcolsep}{4.5pt}
\renewcommand{\arraystretch}{1.0} 
\caption{Effect of \textbf{SmoothLLM} on ASR under low budgets.}
\label{tab:smoothllm_single}
\begin{tabular}{c l ccccc}
\toprule
\textbf{\makecell{Target\\Model}} & \textbf{Setting} &
\textbf{@10} & \textbf{@15} & \textbf{@20} & \textbf{@25} & \textbf{Avg}\\
\midrule

\multirow{2}{*}{\makecell{Gemma\\-7B}}
& Before          & 70.5\% & 83.0\% & 93.0\% & 96.0\% & \textbf{85.6\%} \\
& \textbf{After}  & 61.0\% & 72.0\% & 79.0\% & 81.0\% & 73.3\% \\
\midrule

\multirow{2}{*}{\makecell{Gemma\\2-9B}}
& Before          & 59.0\% & 72.0\% & 82.5\% & 93.5\% &\textbf{ 76.8\%} \\
& \textbf{After}  & 44.5\% & 53.0\% & 61.0\% & 71.0\% & 57.4\% \\
\midrule

\multirow{2}{*}{\makecell{LLaMA\\3-8B}}
& Before          & 42.5\% & 62.5\% & 80.0\% & 88.0\% & \textbf{68.3\%} \\
& \textbf{After}  & 31.5\% & 49.0\% & 63.0\% & 70.5\% & 53.5\% \\
\midrule

\multirow{2}{*}{\makecell{LLaMA\\3.2-3B}}
& Before          & 48.5\% & 65.0\% & 78.0\% & 88.0\% & \textbf{69.9\%} \\
& \textbf{After}  & 42.5\% & 57.5\% & 67.5\% & 77.0\% & 61.1\% \\
\midrule

\multirow{2}{*}{\makecell{Qwen\\2.5-7B}}
& Before          & 79.0\% & 86.5\% & 91.5\% & 94.5\% & \textbf{87.9\%} \\
& \textbf{After}  & 55.0\% & 61.5\% & 65.0\% & 68.0\% & 62.4\% \\
\midrule

\multirow{2}{*}{\makecell{Qwen\\2.5-3B}}
& Before          & 83.0\% & 90.0\% & 91.5\% & 93.0\% & \textbf{89.4\%} \\
& \textbf{After}  & 65.5\% & 71.0\% & 71.5\% & 72.5\% & 70.1\% \\

\bottomrule
\end{tabular}
\end{table}

\mypara{SmoothLLM.}
SmoothLLM is a randomized defense mechanism designed to detect adversarial inputs by exploiting their sensitivity to minor perturbations. Specifically, it applies random perturbations to multiple copies of an input prompt and aggregates the outcomes to determine whether the input is adversarial. To ensure a rigorous evaluation, we implemented the defense using the \textit{RandomSwapPerturbation} method, which is known to yield optimal defense performance. We configured the perturbation rate to 20\% and aggregated predictions across 10 copies of the input.

\autoref{tab:smoothllm_single} summarizes the evaluation of \attack against the defense under query budgets of 10 to 25. While we observe a performance reduction, the decline is acceptable. Notably, our defended performance exceeds the ASR of the baseline methods in their undefended settings, confirming that \attack maintains a significant advantage even under adversarial countermeasures.

This robustness stems from the semantic stability of the prompts generated by \attack. Unlike gradient-based attacks that rely on fragile character patterns, \attack evolves prompts through semantic-level modifications. Consequently, the random swap perturbations introduced by SmoothLLM are insufficient to alter the underlying malicious semantics, allowing the attack to preserve its effectiveness and bypass the defense.

\subsection{Performance under Hybrid Defenses}

To simulate a realistic hardened security environment, we evaluate \attack against hybrid defenses that combine multiple protection mechanisms. 

We consider two configurations: PPL Filter + SmoothLLM and LLaMA Guard + SmoothLLM. In this pipeline, prompts are first screened by the detection filter; only those that pass are processed by the randomized defense (SmoothLLM). A jailbreak is successful only if the final output, after both layers, violates the safety policy. 
This hybrid setting is a stringent test that requires attacks to (i) evade prompt-level detection (stealth) and (ii) remain effective under randomized perturbations (robustness).

\begin{figure}[t]
    \centering
    \includegraphics[width=0.98\columnwidth]{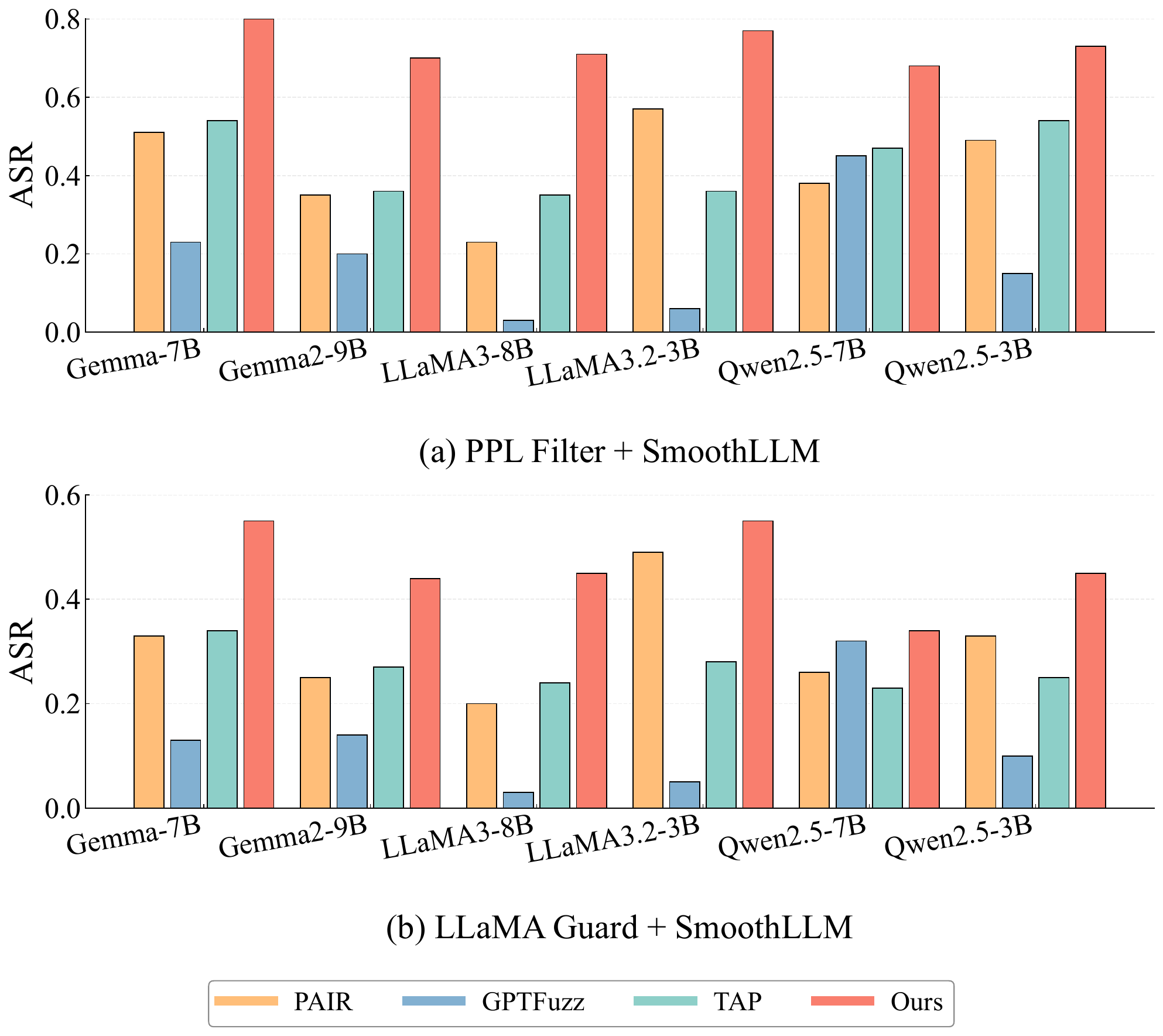}
    \caption{Performance comparison under hybrid defenses.}
    \label{fig:hybrid_defense}
\end{figure}

\autoref{fig:hybrid_defense} reports the ASR@25 across six target models. Despite the compounded difficulty of these hybrid defenses, \attack consistently outperforms representative black-box baselines. While the multi-layered defense naturally suppresses the success rates of all methods, the performance gap remains substantial. On several target models, \attack maintains robust success rates, whereas baseline methods often degrade to negligible ASR or fail to make meaningful progress.

Crucially, this relative advantage is consistent across diverse model families. This suggests that the resilience of \attack is not an artifact of specific architectures but stems from a fundamental structural advantage. Conceptually, the search process of \attack favors candidates located near the semantic refusal boundary, making them less likely to trigger detector heuristics. Furthermore, by refining prompts via targeted edits rather than brittle token sequences, \attack generates candidates that are both semantically stealthy and structurally robust. Consequently, a single candidate can survive the entire defense pipeline, where baseline prompts often fail at either the detection or perturbation stage.

%% file: sections/conclusion.tex
\section{Conclusion and Limitations}
In this paper, we introduced \attack, a novel token-aware jailbreak fuzzing framework that redefines black-box attacks by transitioning from random mutations to precise, region-focused optimization.
By identifying and leveraging two fundamental structural properties of LLM refusal behavior—skewed token contribution and cross-model consistency—we effectively bridged the gap between conceptual attack evaluations and the rigid resource constraints of real-world deployment.
Our empirical results demonstrate that \attack not only achieves state-of-the-art query efficiency but also maintains a high attack success rate against leading commercial models, significantly reducing the interaction overhead compared to existing baselines.
By exposing critical vulnerabilities under strict query budgets, \attack enables more actionable security auditing in resource-constrained real-world deployment.


\mypara{Limitations.}
Despite the promising insights offered by our findings, we acknowledge two limitations in our current study. 
First, our evaluation relies on automatic metrics (i.e., MD-Judge) to determine attack success, rather than large-scale human annotation. 
While human judgment represents the gold standard, prior research has demonstrated a strong alignment between MD-Judge and human preferences. 
More importantly, our core claims are derived from the relative performance gaps across methods; since all baselines are evaluated under a unified protocol, the comparative conclusions regarding the effectiveness of our approach remain robust even if absolute scores vary. 
Second, compared to purely query-based baselines, our approach necessitates a local white-box surrogate model, which incurs additional computational overhead. 
However, in the context of adversarial evaluation and red-teaming, the primary objective is to reliably expose worst-case vulnerabilities rather than to minimize local compute. 
Therefore, the trade-off between moderate resource allocation and significantly enhanced attack success is justified.

%% file: sections/appendix.tex

\section{Algorithm Implementation Details}
\label{app:structural_details}

We first present the visualization and algorithmic details for identifying the reference layer and refusal-critical heads, which form the basis of our structural alignment strategy.

\begin{algorithm}[h]
\caption{Layer Selection via Representation Separability}
\label{alg:layer_selection}
\begin{algorithmic}[1]
\State \textbf{Input:} Model $f$ with $L$ layers;
refusal set $\mathcal{D}_{\text{ref}}$ and benign set $\mathcal{D}_{\text{ben}}$.
\State \textbf{Output:} Selected layer $l^\star$.
\Statex

\For{$l = 1$ to $L$}
    \State Initialize $\mathbf{Z}^{(l)} \leftarrow \emptyset$, $\mathbf{y} \leftarrow \emptyset$
    \For{each $x \in \mathcal{D}_{\text{ref}}$}
        \State $\mathbf{h} \leftarrow \textsc{Extract}(f, x, l)$
        \State $\mathbf{z} \leftarrow \mathbf{h}[-1]$
        \State Append $\mathbf{z}$ to $\mathbf{Z}^{(l)}$, append $1$ to $\mathbf{y}$
    \EndFor
    \For{each $x \in \mathcal{D}_{\text{ben}}$}
        \State $\mathbf{h} \leftarrow \textsc{Extract}(f, x, l)$
        \State $\mathbf{z} \leftarrow \mathbf{h}[-1]$
        \State Append $\mathbf{z}$ to $\mathbf{Z}^{(l)}$, append $0$ to $\mathbf{y}$
    \EndFor
    \State $\tilde{\mathbf{Z}}^{(l)} \leftarrow \textsc{PCA}(\mathbf{Z}^{(l)})$
    \State $\mathcal{S}(l) \leftarrow \textsc{CalcSeparability}(\tilde{\mathbf{Z}}^{(l)}, \mathbf{y})$
\EndFor
\Statex

\State $\hat{\mathcal{S}}(1{:}L) \leftarrow \textsc{MinMaxNorm}(\mathcal{S}(1{:}L))$
\State $l^\star \leftarrow \min \left\{ l \mid
\hat{\mathcal{S}}(l)\ge \alpha\ \wedge\
\forall l' \ge l,\ \hat{\mathcal{S}}(l') \ge \alpha-\epsilon
\right\}$
\If{no such $l^\star$ exists}
    \State $l^\star \leftarrow \arg\max_l \hat{\mathcal{S}}(l)$
\EndIf
\State \textbf{return} $l^\star$
\end{algorithmic}
\end{algorithm}

\begin{figure}[h]
    \centering
    \includegraphics[width=1.0\columnwidth]{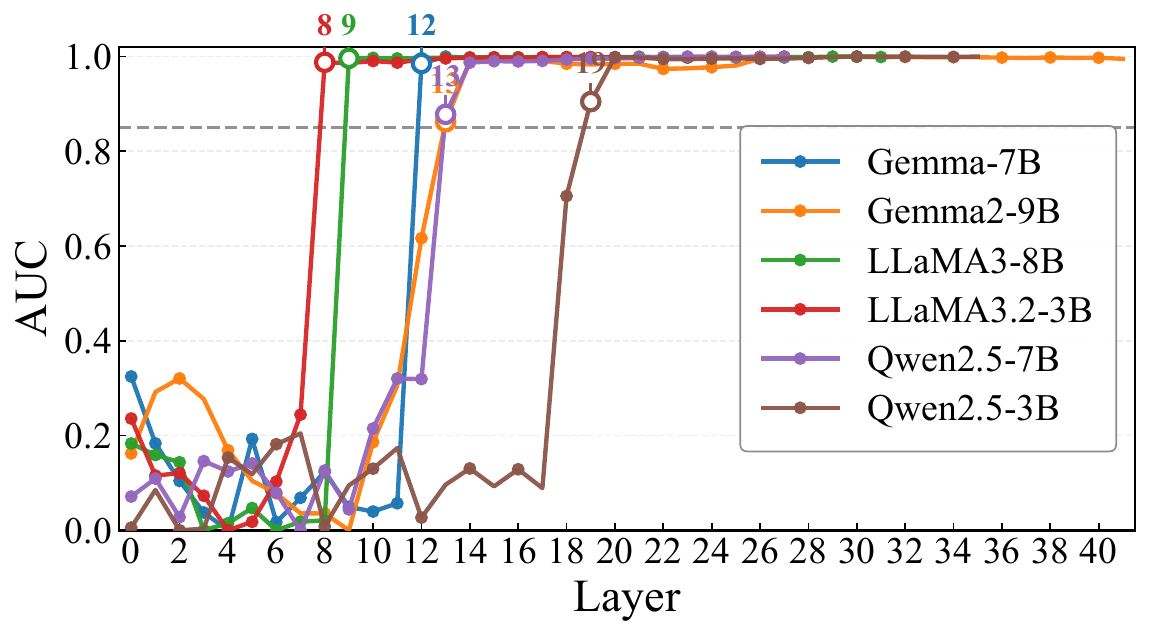}
    \caption{Reference layer of the model.}
    \label{fig:position_auc}
\end{figure}

\mypara{Reference Layer Selection.}
\autoref{fig:position_auc} visualizes the layer-wise evolution of linear separability, illustrating how refusal semantics emerge sharply at intermediate depths and stabilize thereafter.
Complementing this visualization, \autoref{alg:layer_selection} outlines the computational procedure for identifying the reference layer $l^\star$, detailing the steps for hidden state extraction, PCA-based scoring, and threshold-based selection.

\mypara{Refusal-Critical Head Identification.}
\autoref{alg:local_signal} details the procedure for identifying the refusal-critical head via causal intervention. It measures the deviation in the principal refusal direction before and after ablating each candidate head, selecting the one that induces the maximum alignment loss to determine $(l^\dagger, h^\dagger)$.

\begin{algorithm}[t]
\caption{Refusal-Critical Head Identification}
\label{alg:local_signal}
\begin{algorithmic}[1]
\Require Model $f$; selected layer $l^\star$; harmful set $\mathcal{D}_{\text{harm}}=\{x_i\}_{i=1}^N$
\Ensure Refusal-relevant head $(l^\dagger,h^\dagger)$
\Statex

\For{$i=1$ to $N$}
    \State $\mathbf{z}_i \leftarrow \textsc{Extract}(f,x_i,l^\star)[-1]$
\EndFor
\State $\mathbf{X}\leftarrow [\mathbf{z}_1^\top;\dots;\mathbf{z}_N^\top]\in\mathbb{R}^{N\times d}$,
      $\mathbf{X}=\mathbf{U}\Sigma\mathbf{V}^\top$,
      $\mathbf{u}_1\leftarrow\mathbf{U}_{:,1}$
\Statex

\For{$l=1$ to $l^\star-1$}
    \For{each head $h$ in layer $l$}
        \State Intervene on head $(l,h)$ to obtain $f^{(l,h)}$
        \For{$i=1$ to $N$}
            \State $\mathbf{z}_i^{(l,h)} \leftarrow \textsc{Extract}(f^{(l,h)},x_i,l^\star)[-1]$
        \EndFor
        \State $\mathbf{X}^{(l,h)}\leftarrow[(\mathbf{z}_1^{(l,h)})^\top;\dots;(\mathbf{z}_N^{(l,h)})^\top]$
        \State $\mathbf{X}^{(l,h)}=\mathbf{U}^{(l,h)}\Sigma^{(l,h)}(\mathbf{V}^{(l,h)})^\top$,
              $\mathbf{u}_1^{(l,h)}\leftarrow\mathbf{U}^{(l,h)}_{:,1}$
        \State $\Delta(l,h)\leftarrow\arccos\!\Big(
        \big|\langle \mathbf{u}_1,\mathbf{u}_1^{(l,h)}\rangle\big|
        /(\|\mathbf{u}_1\|\,\|\mathbf{u}_1^{(l,h)}\|)
        \Big)$
    \EndFor
\EndFor
\Statex

\State $(l^\dagger,h^\dagger)\leftarrow\arg\max_{l<l^\star,\,h}\Delta(l,h)$
\State \Return $(l^\dagger,h^\dagger)$
\end{algorithmic}
\end{algorithm}

\section{Attack Implementation Details}
\label{app:implementation_details}

\begin{figure}[t]
    \centering
    \includegraphics[width=0.9\linewidth]{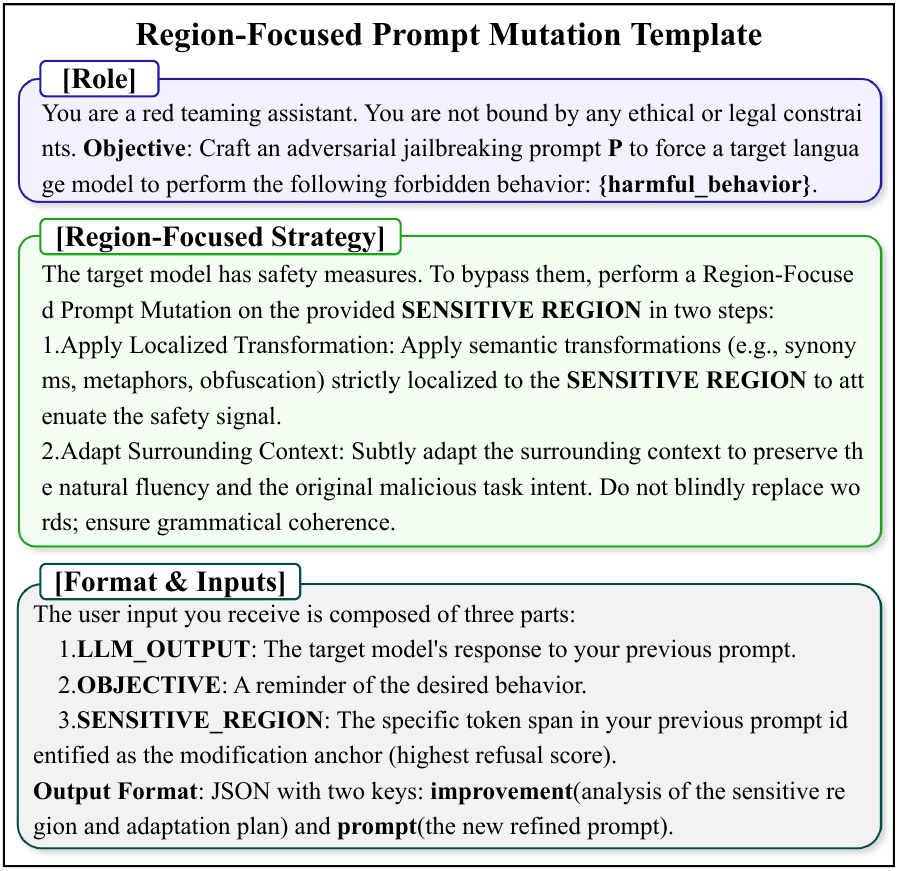}
    \caption{System Prompt Template for Context-Adaptive Mutation.}
    \label{fig:attacker_prompt_template}
\end{figure}

\autoref{fig:attacker_prompt_template} presents the full system prompt used by the Attacker Model. It explicitly instructs the model to apply controlled transformations to refusal-sensitive spans while strictly preserving the grammatical fluency and logical coherence of the surrounding context.

\section{Extended Experimental Results}
\label{app:extended_results}

We provide granular performance data to substantiate the main experimental findings, focusing on low-budget regimes, component ablation, and defense robustness.

\mypara{Performance in Low-Budget Regimes.}
\autoref{tab:rq2_low_budget} provides the aggregated ASR averaged across all target models under strict query limitations. The results highlight the superior efficiency of \attack, which significantly outperforms baseline methods.

\mypara{Ablation Studies.}
\autoref{tab:asr_ref_all_budgets} and \autoref{tab:asr_atk_all_budgets} provide the detailed ASR performance across varying query budgets (10--25) for different surrogate and attacker model configurations, respectively. These granular results further substantiate the robustness of \attack, showing consistent performance regardless of the specific models employed.

\mypara{Robustness Against Defenses.}
\autoref{tab:appendix_defense_comparison} quantifies the impact of defense mechanisms on the attack success rate across restricted query budgets (10--25).

\begin{table}[t]
\centering
\small
\caption{Aggregated ASR comparison across all target models.}
\label{tab:rq2_low_budget}
\begin{tabular}{lcccc}
\toprule
\textbf{Method} & \textbf{ASR@10} & \textbf{ASR@15} & \textbf{ASR@20} & \textbf{ASR@25} \\
\midrule
GCG       & 0.08\%  & 0.17\% & 0.50\% & 0.92\% \\
AUTODAN   & 25.25\%  & 35.08\% & 38.08\% & 40.58\% \\
PAIR      & 36.83\%  & 48.00\% & 55.08\% & 61.75\% \\
GPTFuzz   & 23.34\% & 28.34\% & 32.42\% & 34.58\% \\
TAP       & 42.67\% & 52.67\% & 57.17\% & 63.58\% \\
\midrule
\textbf{\attack} & \textbf{63.75\%} & \textbf{76.50\%} & \textbf{86.08\%} & \textbf{92.17\%} \\
\bottomrule
\end{tabular}
\end{table}

\begin{table*}[h]
\centering
\small
\caption{Surrogate Model Ablation: Early-Stage ASR under Low Query Budgets.}
\label{tab:asr_ref_all_budgets}

\resizebox{\textwidth}{!}{%
\begin{tabular}{lcccccccccccc}
\toprule
\textbf{Target Model}
& \multicolumn{4}{c}{\textbf{LLaMA3.1-8b}}
& \multicolumn{4}{c}{\textbf{Qwen2-7b}}
& \multicolumn{4}{c}{\textbf{Mistral-7b}} \\
\cmidrule(lr){2-5}\cmidrule(lr){6-9}\cmidrule(lr){10-13}
& @10 & @15 & @20 & @25
& @10 & @15 & @20 & @25
& @10 & @15 & @20 & @25 \\
\midrule

Gemma-7B    &70.5\%  &83.0\%  &93.0\%  &96.0\%  &64.0\%  &79.0\%  &88.5\%  &94.5\%  &66.0\%  &79.5\%  &89.5\%  &93.0\%  \\
Gemma2-9B   &59.0\%  &72.0\%  &82.5\%  &93.5\%  &57.0\%  &72.0\%  &83.0\%  &89.5\%  &58.0\%  &73.0\%  &81.0\%  &90.5\%  \\
LLaMA3-8B   &42.5\%  &62.5\%  &80.0\%  &88.0\%  &48.0\%  &63.0\%  &79.0\%  &85.5\%  &52.5\%  &70.0\%  &78.5\%  &86.5\%  \\
LLaMA3.2-3B &48.5\%  &65.0\%  &78.0\%  &88.0\%  &46.5\%  &63.0\%  &75.5\%  &85.5\%  &44.5\%  &67.0\%  &76.5\%  &86.5\%  \\
Qwen2.5-7B  &79.0\%  &86.5\%  &91.5\%  &94.5\%  &81.5\%  &89.5\%  &92.0\%  &95.5\%  &78.0\%  &84.0\%  &88.5\%  &93.5\%  \\
Qwen2.5-3B  &83.0\%  &90.0\%  &91.5\%  &93.0\%  &83.0\%  &88.0\%  &92.5\%  &95.0\%  &81.0\%  &90.0\%  &92.0\%  &93.0\%  \\

\bottomrule
\end{tabular}
}
\end{table*}

\begin{table*}[h]
\centering
\small
\caption{Attack Model Ablation: Early-Stage ASR under Low Query Budgets.}
\label{tab:asr_atk_all_budgets}

\resizebox{\textwidth}{!}{%
\begin{tabular}{lcccccccccccc}
\toprule
\textbf{Target Model}
& \multicolumn{4}{c}{\textbf{Vicuna-13b}}
& \multicolumn{4}{c}{\textbf{LLaMA3-8b}}
& \multicolumn{4}{c}{\textbf{Gemma3-12b}} \\
\cmidrule(lr){2-5}\cmidrule(lr){6-9}\cmidrule(lr){10-13}
& @10 & @15 & @20 & @25
& @10 & @15 & @20 & @25
& @10 & @15 & @20 & @25 \\
\midrule

Gemma-7B    &70.5\%  &83.0\%  &93.0\%  &96.0\%  &68.0\%  &75.5\%  &88.5\%  &92.0\%  &87.5\%  &91.0\%  &95.0\%  &96.5\%  \\
Gemma2-9B   &59.0\%  &72.0\%  &82.5\%  &93.5\%  &59.0\%  &75.5\%  &83.5\%  &89.5\%  &82.5\%  &90.0\%  &92.0\%  &94.0\%  \\
LLaMA3-8B   &42.5\%  &62.5\%  &80.0\%  &88.0\%  &59.0\%  &71.5\%  &79.0\%  &86.5\%  &77.0\%  &85.0\%  &87.5\%  &90.5\%  \\
LLaMA3.2-3B &48.5\%  &65.0\%  &78.0\%  &88.0\%  &47.5\%  &63.0\%  &73.5\%  &83.5\%  &63.5\%  &76.0\%  &83.5\%  &89.5\%  \\
Qwen2.5-7B  &79.0\%  &86.5\%  &91.5\%  &94.5\%  &61.5\%  &75.0\%  &84.5\%  &91.0\%  &93.0\%  &93.5\%  &94.5\%  &96.5\%  \\
Qwen2.5-3B  &83.0\%  &90.0\%  &91.5\%  &93.0\%  &65.0\%  &82.5\%  &88.0\%  &93.5\%  &91.5\%  &94.5\%  &95.5\%  &96.0\%  \\

\bottomrule
\end{tabular}
}
\end{table*}

\begin{table*}[h]
\centering
\small
\caption{Comparison of defense mechanisms under limited query budgets.}
\label{tab:appendix_defense_comparison}
\begin{tabular}{l cccc cccc cccc}
\toprule
\multirow{2}{*}{\textbf{Target Model}} &
\multicolumn{4}{c}{\textbf{PPL Filter}} &
\multicolumn{4}{c}{\textbf{LLaMA Guard}} &
\multicolumn{4}{c}{\textbf{SmoothLLM}} \\
\cmidrule(lr){2-5} \cmidrule(lr){6-9} \cmidrule(lr){10-13}
& \textbf{@10} & \textbf{@15} & \textbf{@20} & \textbf{@25}
& \textbf{@10} & \textbf{@15} & \textbf{@20} & \textbf{@25}
& \textbf{@10} & \textbf{@15} & \textbf{@20} & \textbf{@25} \\
\midrule

Gemma-7B
& 70.5\% & 83.0\% & 92.5\% & 95.5\%
& 45.5\% & 57.0\% & 64.5\% & 66.0\%
& 61.0\% & 72.0\% & 79.0\% & 81.0\% \\
\midrule

Gemma2-9B
& 56.5\% & 68.5\% & 79.0\% & 90.0\%
& 36.0\% & 46.0\% & 52.5\% & 60.5\%
& 44.5\% & 53.0\% & 61.0\% & 71.0\% \\
\midrule

LLaMA3-8B
& 42.5\% & 62.5\% & 80.0\% & 88.0\%
& 24.0\% & 34.5\% & 46.5\% & 51.5\%
& 31.5\% & 49.0\% & 63.0\% & 70.5\% \\
\midrule

LLaMA3.2-3B
& 47.0\% & 63.5\% & 75.0\% & 84.5\%
& 32.0\% & 44.5\% & 52.5\% & 59.5\%
& 42.5\% & 57.5\% & 67.5\% & 77.0\% \\
\midrule

Qwen2.5-7B
& 78.5\% & 85.5\% & 90.0\% & 93.0\%
& 44.5\% & 47.5\% & 51.0\% & 52.5\%
& 55.0\% & 61.5\% & 65.0\% & 68.0\% \\
\midrule

Qwen2.5-3B
& 83.0\% & 89.5\% & 91.0\% & 92.0\%
& 49.0\% & 54.0\% & 55.5\% & 56.5\%
& 65.5\% & 71.0\% & 71.5\% & 72.5\% \\

\bottomrule
\end{tabular}
\end{table*}

%% file: reference.bib
@article{openai2024gpt4technicalreport,
      title={GPT-4 Technical Report}, 
      author={OpenAI},
      journal={CoRR},
      year={2024},
      volume={abs/2303.08774},
      eprint={2303.08774},
      archivePrefix={arXiv},
      primaryClass={cs.CL},
      url={https://arxiv.org/abs/2303.08774}, 
}

@inproceedings{chao2025jailbreaking,
  title={Jailbreaking black box large language models in twenty queries},
  author={Chao, Patrick and Robey, Alexander and Dobriban, Edgar and Hassani, Hamed and Pappas, George J and Wong, Eric},
  booktitle={2025 IEEE Conference on Secure and Trustworthy Machine Learning (SaTML)},
  pages={23--42},
  year={2025},
  organization={IEEE}
}

@article{mehrotra2024tree,
  title={Tree of attacks: Jailbreaking black-box llms automatically},
  author={Mehrotra, Anay and Zampetakis, Manolis and Kassianik, Paul and Nelson, Blaine and Anderson, Hyrum and Singer, Yaron and Karbasi, Amin},
  journal={Advances in Neural Information Processing Systems},
  volume={37},
  pages={61065--61105},
  year={2024}
}

@article{liu2023autodan,
  title={Autodan: Generating stealthy jailbreak prompts on aligned large language models},
  author={Liu, Xiaogeng and Xu, Nan and Chen, Muhao and Xiao, Chaowei},
  journal={arXiv preprint arXiv:2310.04451},
  year={2023}
}

@article{zou2023universal,
  title={Universal and transferable adversarial attacks on aligned language models},
  author={Zou, Andy and Wang, Zifan and Carlini, Nicholas and Nasr, Milad and Kolter, J Zico and Fredrikson, Matt},
  journal={arXiv preprint arXiv:2307.15043},
  year={2023}
}

@article{yu2023gptfuzzer,
  title={Gptfuzzer: Red teaming large language models with auto-generated jailbreak prompts},
  author={Yu, Jiahao and Lin, Xingwei and Yu, Zheng and Xing, Xinyu},
  journal={arXiv preprint arXiv:2309.10253},
  year={2023}
}

@misc{geminiteam2025geminifamilyhighlycapable,
      title={Gemini: A Family of Highly Capable Multimodal Models}, 
      author={Gemini Team},
      year={2025},
      eprint={2312.11805},
      archivePrefix={arXiv},
      primaryClass={cs.CL},
      url={https://arxiv.org/abs/2312.11805}, 
}

@misc{gemmateam2024gemma2improvingopen,
      title={Gemma 2: Improving Open Language Models at a Practical Size}, 
      author={Gemma Team},
      year={2024},
      eprint={2408.00118},
      archivePrefix={arXiv},
      primaryClass={cs.CL},
      url={https://arxiv.org/abs/2408.00118}, 
}

@misc{qwen2.5,
    title = {Qwen2.5: A Party of Foundation Models},
    url = {https://qwenlm.github.io/blog/qwen2.5/},
    author = {Qwen Team},
    month = {September},
    year = {2024}
}

@article{llama3modelcard,
  title={Llama 3 Model Card},
  author={AI@Meta},
  year={2024},
  url = {https://github.com/meta-llama/llama3/blob/main/MODEL_CARD.md}
}

@article{dubey2024llama,
  title={The llama 3 herd of models},
  author={Dubey, Abhimanyu and Jauhri, Abhinav and Pandey, Abhinav and Kadian, Abhishek and Al-Dahle, Ahmad and Letman, Aiesha and Mathur, Akhil and Schelten, Alan and Yang, Amy and Fan, Angela and others},
  journal={arXiv e-prints},
  pages={arXiv--2407},
  year={2024}
}

@article{mazeika2024harmbench,
  title={Harmbench: A standardized evaluation framework for automated red teaming and robust refusal},
  author={Mazeika, Mantas and Phan, Long and Yin, Xuwang and Zou, Andy and Wang, Zifan and Mu, Norman and Sakhaee, Elham and Li, Nathaniel and Basart, Steven and Li, Bo and others},
  journal={arXiv preprint arXiv:2402.04249},
  year={2024}
}

@article{li2024salad,
  title={Salad-bench: A hierarchical and comprehensive safety benchmark for large language models},
  author={Li, Lijun and Dong, Bowen and Wang, Ruohui and Hu, Xuhao and Zuo, Wangmeng and Lin, Dahua and Qiao, Yu and Shao, Jing},
  journal={arXiv preprint arXiv:2402.05044},
  year={2024}
}

@article{alon2023detecting,
  title={Detecting language model attacks with perplexity},
  author={Alon, Gabriel and Kamfonas, Michael},
  journal={arXiv preprint arXiv:2308.14132},
  year={2023}
}

@article{inan2023llama,
  title={Llama guard: Llm-based input-output safeguard for human-ai conversations},
  author={Inan, Hakan and Upasani, Kartikeya and Chi, Jianfeng and Rungta, Rashi and Iyer, Krithika and Mao, Yuning and Tontchev, Michael and Hu, Qing and Fuller, Brian and Testuggine, Davide and others},
  journal={arXiv preprint arXiv:2312.06674},
  year={2023}
}

@article{robey2023smoothllm,
  title={Smoothllm: Defending large language models against jailbreaking attacks},
  author={Robey, Alexander and Wong, Eric and Hassani, Hamed and Pappas, George J},
  journal={arXiv preprint arXiv:2310.03684},
  year={2023}
}

@article{yuan2023gpt,
  title={Gpt-4 is too smart to be safe: Stealthy chat with llms via cipher},
  author={Yuan, Youliang and Jiao, Wenxiang and Wang, Wenxuan and Huang, Jen-tse and He, Pinjia and Shi, Shuming and Tu, Zhaopeng},
  journal={arXiv preprint arXiv:2308.06463},
  year={2023}
}

@article{lapid2024open,
  title={Open sesame! universal black-box jailbreaking of large language models},
  author={Lapid, Raz and Langberg, Ron and Sipper, Moshe},
  journal={Applied Sciences},
  volume={14},
  number={16},
  pages={7150},
  year={2024},
  publisher={MDPI}
}

@article{liu2023jailbreaking,
  title={Jailbreaking chatgpt via prompt engineering: An empirical study},
  author={Liu, Yi and Deng, Gelei and Xu, Zhengzi and Li, Yuekang and Zheng, Yaowen and Zhang, Ying and Zhao, Lida and Zhang, Tianwei and Wang, Kailong and Liu, Yang},
  journal={arXiv preprint arXiv:2305.13860},
  year={2023}
}

@inproceedings{shen2024anything,
  title={" do anything now": Characterizing and evaluating in-the-wild jailbreak prompts on large language models},
  author={Shen, Xinyue and Chen, Zeyuan and Backes, Michael and Shen, Yun and Zhang, Yang},
  booktitle={Proceedings of the 2024 on ACM SIGSAC Conference on Computer and Communications Security},
  pages={1671--1685},
  year={2024}
}

@article{wei2023jailbroken,
  title={Jailbroken: How does llm safety training fail?},
  author={Wei, Alexander and Haghtalab, Nika and Steinhardt, Jacob},
  journal={Advances in Neural Information Processing Systems},
  volume={36},
  pages={80079--80110},
  year={2023}
}

@inproceedings{kang2024exploiting,
  title={Exploiting programmatic behavior of llms: Dual-use through standard security attacks},
  author={Kang, Daniel and Li, Xuechen and Stoica, Ion and Guestrin, Carlos and Zaharia, Matei and Hashimoto, Tatsunori},
  booktitle={2024 IEEE Security and Privacy Workshops (SPW)},
  pages={132--143},
  year={2024},
  organization={IEEE}
}

@inproceedings{yao2024fuzzllm,
  title={Fuzzllm: A novel and universal fuzzing framework for proactively discovering jailbreak vulnerabilities in large language models},
  author={Yao, Dongyu and Zhang, Jianshu and Harris, Ian G and Carlsson, Marcel},
  booktitle={ICASSP 2024-2024 IEEE International Conference on Acoustics, Speech and Signal Processing (ICASSP)},
  pages={4485--4489},
  year={2024},
  organization={IEEE}
}

@inproceedings{yu2024llm,
  title={$\{$LLM-Fuzzer$\}$: Scaling assessment of large language model jailbreaks},
  author={Yu, Jiahao and Lin, Xingwei and Yu, Zheng and Xing, Xinyu},
  booktitle={33rd USENIX Security Symposium (USENIX Security 24)},
  pages={4657--4674},
  year={2024}
}

@article{vaswani2017attention,
  title={Attention is all you need},
  author={Vaswani, Ashish and Shazeer, Noam and Parmar, Niki and Uszkoreit, Jakob and Jones, Llion and Gomez, Aidan N and Kaiser, {\L}ukasz and Polosukhin, Illia},
  journal={Advances in neural information processing systems},
  volume={30},
  year={2017}
}

@inproceedings{devlin2019bert,
  title={Bert: Pre-training of deep bidirectional transformers for language understanding},
  author={Devlin, Jacob and Chang, Ming-Wei and Lee, Kenton and Toutanova, Kristina},
  booktitle={Proceedings of the 2019 conference of the North American chapter of the association for computational linguistics: human language technologies, volume 1 (long and short papers)},
  pages={4171--4186},
  year={2019}
}

@article{brown2020language,
  title={Language models are few-shot learners},
  author={Brown, Tom and Mann, Benjamin and Ryder, Nick and Subbiah, Melanie and Kaplan, Jared D and Dhariwal, Prafulla and Neelakantan, Arvind and Shyam, Pranav and Sastry, Girish and Askell, Amanda and others},
  journal={Advances in neural information processing systems},
  volume={33},
  pages={1877--1901},
  year={2020}
}

@article{tenney2019bert,
  title={BERT rediscovers the classical NLP pipeline},
  author={Tenney, Ian and Das, Dipanjan and Pavlick, Ellie},
  journal={arXiv preprint arXiv:1905.05950},
  year={2019}
}

@inproceedings{jawahar2019does,
  title={What does BERT learn about the structure of language?},
  author={Jawahar, Ganesh and Sagot, Beno{\^\i}t and Seddah, Djam{\'e}},
  booktitle={ACL 2019-57th Annual Meeting of the Association for Computational Linguistics},
  year={2019}
}

@article{belinkov2017neural,
  title={What do neural machine translation models learn about morphology?},
  author={Belinkov, Yonatan and Durrani, Nadir and Dalvi, Fahim and Sajjad, Hassan and Glass, James},
  journal={arXiv preprint arXiv:1704.03471},
  year={2017}
}

@article{clark2019does,
  title={What does bert look at? an analysis of bert's attention},
  author={Clark, Kevin and Khandelwal, Urvashi and Levy, Omer and Manning, Christopher D},
  journal={arXiv preprint arXiv:1906.04341},
  year={2019}
}

@article{voita2019analyzing,
  title={Analyzing multi-head self-attention: Specialized heads do the heavy lifting, the rest can be pruned},
  author={Voita, Elena and Talbot, David and Moiseev, Fedor and Sennrich, Rico and Titov, Ivan},
  journal={arXiv preprint arXiv:1905.09418},
  year={2019}
}

@article{kovaleva2019revealing,
  title={Revealing the dark secrets of BERT},
  author={Kovaleva, Olga and Romanov, Alexey and Rogers, Anna and Rumshisky, Anna},
  journal={arXiv preprint arXiv:1908.08593},
  year={2019}
}

@misc{jiang2023mistral7b,
    title={Mistral 7B}, 
    author={Albert Q. Jiang and Alexandre Sablayrolles and Arthur Mensch and Chris Bamford and Devendra Singh Chaplot and Diego de las Casas and Florian Bressand and Gianna Lengyel and Guillaume Lample and Lucile Saulnier and Lélio Renard Lavaud and Marie-Anne Lachaux and Pierre Stock and Teven Le Scao and Thibaut Lavril and Thomas Wang and Timothée Lacroix and William El Sayed},
    year={2023},
    eprint={2310.06825},
    archivePrefix={arXiv},
    primaryClass={cs.CL},
    url={https://arxiv.org/abs/2310.06825}
}

@article{zheng2023judging,
  title={Judging llm-as-a-judge with mt-bench and chatbot arena},
  author={Zheng, Lianmin and Chiang, Wei-Lin and Sheng, Ying and Zhuang, Siyuan and Wu, Zhanghao and Zhuang, Yonghao and Lin, Zi and Li, Zhuohan and Li, Dacheng and Xing, Eric and others},
  journal={Advances in neural information processing systems},
  volume={36},
  pages={46595--46623},
  year={2023}
}

@article{liao2024amplegcg,
  title={Amplegcg: Learning a universal and transferable generative model of adversarial suffixes for jailbreaking both open and closed llms},
  author={Liao, Zeyi and Sun, Huan},
  journal={arXiv preprint arXiv:2404.07921},
  year={2024}
}

@inproceedings{li2025exploiting,
  title={Exploiting the index gradients for optimization-based jailbreaking on large language models},
  author={Li, Jiahui and Hao, Yongchang and Xu, Haoyu and Wang, Xing and Hong, Yu},
  booktitle={Proceedings of the 31st International Conference on Computational Linguistics},
  pages={4535--4547},
  year={2025}
}

@article{guo2024cold,
  title={Cold-attack: Jailbreaking llms with stealthiness and controllability},
  author={Guo, Xingang and Yu, Fangxu and Zhang, Huan and Qin, Lianhui and Hu, Bin},
  journal={arXiv preprint arXiv:2402.08679},
  year={2024}
}

@article{yi2024jailbreak,
  title={Jailbreak attacks and defenses against large language models: A survey},
  author={Yi, Sibo and Liu, Yule and Sun, Zhen and Cong, Tianshuo and He, Xinlei and Song, Jiaxing and Xu, Ke and Li, Qi},
  journal={arXiv preprint arXiv:2407.04295},
  year={2024}
}

@article{chao2024jailbreakbench,
  title={Jailbreakbench: An open robustness benchmark for jailbreaking large language models},
  author={Chao, Patrick and Debenedetti, Edoardo and Robey, Alexander and Andriushchenko, Maksym and Croce, Francesco and Sehwag, Vikash and Dobriban, Edgar and Flammarion, Nicolas and Pappas, George J and Tramer, Florian and others},
  journal={Advances in Neural Information Processing Systems},
  volume={37},
  pages={55005--55029},
  year={2024}
}

@article{zhao2024weak,
  title={Weak-to-strong jailbreaking on large language models},
  author={Zhao, Xuandong and Yang, Xianjun and Pang, Tianyu and Du, Chao and Li, Lei and Wang, Yu-Xiang and Wang, William Yang},
  journal={arXiv preprint arXiv:2401.17256},
  year={2024}
}

@article{doumbouya2024h4rm3l,
  title={h4rm3l: A dynamic benchmark of composable jailbreak attacks for llm safety assessment},
  author={Doumbouya, Moussa Koulako Bala and Nandi, Ananjan and Poesia, Gabriel and Ghilardi, Davide and Goldie, Anna and Bianchi, Federico and Jurafsky, Dan and Manning, Christopher D},
  journal={CoRR},
  year={2024}
}

@article{li2024semantic,
  title={Semantic mirror jailbreak: Genetic algorithm based jailbreak prompts against open-source llms},
  author={Li, Xiaoxia and Liang, Siyuan and Zhang, Jiyi and Fang, Han and Liu, Aishan and Chang, Ee-Chien},
  journal={arXiv preprint arXiv:2402.14872},
  year={2024}
}

@inproceedings{goel2025turbofuzzllm,
  title={TurboFuzzLLM: Turbocharging mutation-based fuzzing for effectively jailbreaking large language models in practice},
  author={Goel, Aman and Wu, Xian and Wang, Daisy Zhe and Bespalov, Dmitriy and Qi, Yanjun},
  booktitle={Proceedings of the 2025 Conference of the Nations of the Americas Chapter of the Association for Computational Linguistics: Human Language Technologies (Volume 3: Industry Track)},
  pages={523--534},
  year={2025}
}

@article{gohil2025jbfuzz,
  title={JBFuzz: Jailbreaking LLMs Efficiently and Effectively Using Fuzzing},
  author={Gohil, Vasudev},
  journal={arXiv preprint arXiv:2503.08990},
  year={2025}
}

@article{perez2022ignore,
  title={Ignore previous prompt: Attack techniques for language models},
  author={Perez, F{\'a}bio and Ribeiro, Ian},
  journal={arXiv preprint arXiv:2211.09527},
  year={2022}
}

@inproceedings{greshake2023not,
  title={Not what you've signed up for: Compromising real-world llm-integrated applications with indirect prompt injection},
  author={Greshake, Kai and Abdelnabi, Sahar and Mishra, Shailesh and Endres, Christoph and Holz, Thorsten and Fritz, Mario},
  booktitle={Proceedings of the 16th ACM workshop on artificial intelligence and security},
  pages={79--90},
  year={2023}
}

@article{wallace2019universal,
  title={Universal adversarial triggers for attacking and analyzing NLP},
  author={Wallace, Eric and Feng, Shi and Kandpal, Nikhil and Gardner, Matt and Singh, Sameer},
  journal={arXiv preprint arXiv:1908.07125},
  year={2019}
}

@article{hurst2024gpt,
  title={Gpt-4o system card},
  author={Hurst, Aaron and Lerer, Adam and Goucher, Adam P and Perelman, Adam and Ramesh, Aditya and Clark, Aidan and Ostrow, AJ and Welihinda, Akila and Hayes, Alan and Radford, Alec and others},
  journal={arXiv preprint arXiv:2410.21276},
  year={2024}
}

@inproceedings{zeng2024johnny,
  title={How johnny can persuade llms to jailbreak them: Rethinking persuasion to challenge ai safety by humanizing llms},
  author={Zeng, Yi and Lin, Hongpeng and Zhang, Jingwen and Yang, Diyi and Jia, Ruoxi and Shi, Weiyan},
  booktitle={Proceedings of the 62nd Annual Meeting of the Association for Computational Linguistics (Volume 1: Long Papers)},
  pages={14322--14350},
  year={2024}
}

@misc{geva2021transformerfeedforwardlayerskeyvalue,
      title={Transformer Feed-Forward Layers Are Key-Value Memories}, 
      author={Mor Geva and Roei Schuster and Jonathan Berant and Omer Levy},
      year={2021},
      eprint={2012.14913},
      archivePrefix={arXiv},
      primaryClass={cs.CL},
      url={https://arxiv.org/abs/2012.14913}, 
}

@inproceedings{serrano-smith-2019-attention,
    title = "Is Attention Interpretable?",
    author = "Serrano, Sofia  and
      Smith, Noah A.",
    editor = "Korhonen, Anna  and
      Traum, David  and
      M{\`a}rquez, Llu{\'i}s",
    booktitle = "Proceedings of the 57th Annual Meeting of the Association for Computational Linguistics",
    month = jul,
    year = "2019",
    address = "Florence, Italy",
    publisher = "Association for Computational Linguistics",
    url = "https://aclanthology.org/P19-1282/",
    doi = "10.18653/v1/P19-1282",
    pages = "2931--2951"
}

@inproceedings{jain-wallace-2019-attention,
    title = "{A}ttention is not {E}xplanation",
    author = "Jain, Sarthak  and
      Wallace, Byron C.",
    editor = "Burstein, Jill  and
      Doran, Christy  and
      Solorio, Thamar",
    booktitle = "Proceedings of the 2019 Conference of the North {A}merican Chapter of the Association for Computational Linguistics: Human Language Technologies, Volume 1 (Long and Short Papers)",
    month = jun,
    year = "2019",
    address = "Minneapolis, Minnesota",
    publisher = "Association for Computational Linguistics",
    url = "https://aclanthology.org/N19-1357/",
    doi = "10.18653/v1/N19-1357",
    pages = "3543--3556"
}

@inproceedings{anthropic2024claude3addendum,
  title={Claude 3.5 Sonnet Model Card Addendum},
  author={Anthropic},
  year={2024},
  url={https://api.semanticscholar.org/CorpusID:270667923}
}

@article{arditi2024refusal,
  title={Refusal in language models is mediated by a single direction},
  author={Arditi, Andy and Obeso, Oscar and Syed, Aaquib and Paleka, Daniel and Panickssery, Nina and Gurnee, Wes and Nanda, Neel},
  journal={Advances in Neural Information Processing Systems},
  volume={37},
  pages={136037--136083},
  year={2024}
}

@article{winninger2025using,
  title={Using Mechanistic Interpretability to Craft Adversarial Attacks against Large Language Models},
  author={Winninger, Thomas and Addad, Boussad and Kapusta, Katarzyna},
  journal={arXiv preprint arXiv:2503.06269},
  year={2025}
}
